\documentclass[useAMS,usenatbib]{mn2e}
\usepackage{amsmath}
\usepackage{graphicx}
\usepackage{epstopdf}
\usepackage{amssymb}
\usepackage{extarrows}
\usepackage{bm}
\usepackage{float}
\usepackage{color}
\newcommand{\ba}{\begin{eqnarray}}
\newcommand{\ea}{\end{eqnarray}}
\newcommand{\be}{\begin{equation}}
\newcommand{\ee}{\end{equation}}
\newcommand{\gr}{\mathrm{GR}}

\newcommand{\m}{\mathrm{max}}
\newcommand{\mi}{\mathrm{min}}

\newcommand{\oct}{\mathrm{oct}}

\newcommand{\au}{\mathrm{AU}}

\newcommand{\IN}{\mathrm{in}}
\newcommand{\OUT}{\mathrm{out}}

\newcommand{\lk}{\mathrm{LK}}
\newcommand{\merger}{\mathrm{merger}}
\newcommand{\eff}{\mathrm{eff}}

\newcommand{\li}{\mathrm{lim}}
\newcommand{\tot}{\mathrm{tot}}

\newcommand{\uvec}{\hat{\mathbf{e}}}
\newcommand{\nvec}{\hat{\mathbf{L}}}
\newcommand{\jvec}{\mathbf{j}}
\newcommand{\evec}{\mathbf{e}}

\def\e1{e_1^2}

\title[Enhanced Black Hole Mergers in Binary-Binary Interactions]{Enhanced Black-Hole Mergers in Binary-Binary Interactions}

\author[Liu, \& Lai]
{Bin Liu$^{1,2}$, Dong Lai$^{1,2,3}$\\
$^{1}$ Cornell Center for Astrophysics and Planetary Science, Cornell University, Ithaca, NY 14853, USA\\
$^{2}$ Shanghai Astronomical Observatory, Chinese Academy of Sciences, 80Nandan Road, Shanghai 200030, China\\
$^{3}$ Tsung-Dao Lee Institute, Shanghai 200240, China
}

\begin{document}


\pagerange{\pageref{firstpage}--\pageref{lastpage}} \pubyear{2018}

\maketitle

\label{firstpage}

\begin{abstract}
We study the orbital evolution of black hole (BH) binaries in
quadruple systems, where the tertiary binary excites large
eccentricity in the BH binary through Lidov-Kozai (LK) oscillations,
causing the binary BHs to merge via gravitational radiation.
For typical BH binaries with masses $m_{1,2}\simeq 20M_\odot-30M_\odot$ and
initial semimajor axis $a_0\sim100$ AU
(such that the binaries have no chance of merging by themselves within $\sim10^{10}$ yrs),
we show that binary-binary interactions can significantly increase the LK
window for mergers (the range of companion inclinations that allows the BH binary
to merge within 10~Gyrs). This increase arises from a secular
resonance between the LK oscillation of the BH binary and the nodal
precession of the outer (binary-binary) orbit driven by the tertiary
binary. Therefore, in the presence of tertiary binary, the BH merger
fraction is increased to $10-30\%$, an order of magnitude larger than
the merger fraction found in similar triple systems.
Our analysis (with appropriate scalings) can be easily adapted to other configurations of systems, such as relatively compact
BH binaries and moderately hierarchical triples, which may generate even higher merger fractions.
Since the occurrence rate of stellar quadruples in the galactic fields is not
much smaller than that of stellar triples, our result suggests that
dynamically induced BH mergers in quadruple systems may be an
important channel of producing BH mergers observed by LIGO/VIRGO.
\end{abstract}


\section{Introduction}
\label{sec 1}

Since 2015, a number of black hole (BH) binary and
neutron star (NS) binary mergers have been observed in gravitational
waves by aLIGO/VIRGO \citep[e.g.,][]{Abbott 2016a,Abbott
  2016b,Abbott 2017a,Abbott 2017b,Abbott 2017c,Abbott 2017d}.
To bring two BHs into sufficiently close orbits and allow
gravitational-radiation driven binary coalescence,
several different formation scenarios have been proposed.
These include isolated binary evolution,
either through common-envelop phases
\citep[e.g.,][]{Lipunov 1997,Lipunov 2017,Podsiadlowski
  2003,Belczynski 2010,Belczynski 2016,Dominik 2012, Dominik
  2013,Dominik 2015} or through chemically homogeneous evolution
associated with rapid stellar rotations \citep[e.g.,][]{Mandel and de
  Mink 2016,Marchant 2016}, three-body encounters and/or secular interactions
in dense star clusters such as globular cluster \citep[e.g.,][]{Portegies
  2000,Miller 2002,Wen 2003,Miller 2009,O'Leary 2006, Banerjee 2010,Downing
  2010,Thompson 2011,Rodriguez 2015,Chatterjee 2017,Samsing 2018} or galactic nuclei
\citep[e.g.,][]{O'Leary 2009,Antonini 2012,Antonini and Rasio
  2016,VanLandingham 2016,Petrovich 2017,Hoang 2017,Leigh 2018}
, and secular/nonsecular Lidov-Kozai oscillations \citep[e.g.,][]{Lidov,Kozai,Naoz 2016} in isolated triples in the galactic fields
\citep[e.g.,][]{Silsbee and Tremaine 2017,Antonini 2017,Liu-ApJ}.

The BH binary merger rate inferred from the LIGO detections (10-200 Gpc$^{-3}$yr$^{-1}$)
is higher than expected and challenges existing models.
Additional mechanisms/effects may be required to produce a greater BH merger rate to match observations.
Lidov-Kozai (LK) oscillations driven by tertiary companions
(either another star/BH in the galactic triple scenario, or a supermassive BH for binaries near galactic nuclei)
provide a natural, purely dynamical mechanism to induce binary BH merger
\citep[e.g.,][]{Miller 2002,Wen 2003,Thompson 2011,Antonini 2012,Antonini 2014,Hoang 2017}.
In a recent paper \citep[][]{Liu-ApJ}, we systematically study the merger window
(the range of companion inclinations that allows the inner binary to merge within $\sim$10~Gyrs)
and merger fraction for BH binaries in triples for a wide range of parameters,
taking account of both (octupole-level) secular and non-secular effects.
We find that for a ``typical" inner binary system
(with masses $m_1=30M_\odot$, $m_2=20M_\odot$, initial separation $a_\IN=100$ AU)
\footnote{
Note that in the isolated binary
evolution scenario, the massive stellar binaries must be formed with relatively
small separations ($\lesssim 10$~AU), so that binary interactions (mass transfer and common envelop evolution) can shrink the orbit \citep[e.g.,][]{Belczynski 2016}.
Thus, we consider BH binaries with initially wider separations that
cannot merge within the age of universe by themselves.
}\label{fn:1}
and a
random orientation of the tertiary binary orbits, the merger fraction ranges from
$\sim 1\%$ at $e_{\rm out}=0$ (quadrupole LK effect) to $\sim10-20\%$ at $e_{\rm out}=0.9$ (octupole LK effect).

The merger fraction of BH binaries in triples can increases when the tertiary companion
is a binary by itself (see Figure \ref{fig:configuration}).
Such binary-binary systems may allow Lidov-Kozai (LK) eccentricity excitation to operate over a
wide range of inclinations \citep[e.g.,][]{Pejcha quadple,Vokrouhlicky quadruple,Hamers 2016}.
The qualitative reason is as follows \citep[][]{Hamers and Lai 2017}:
the second binary induces nodal precession of the outer binary (at the characteristic rate $\Omega_\OUT$);
when $\Omega_\OUT$ matches the LK rate of the first (inner) binary, a secular resonance occurs;
this can generate large mutual inclinations (between the first binary and the outer binary),
and therefore induce eccentricity excitation of the first (inner) binary.
\citet{Fang 2018} and \citet{Hamers (2018)} studied this ``enhanced LK effect" in the context of
white dwarf (WD) binaries, with emphasis on WD-WD mergers relevant to Type Ia supernovae.
\citet{Petrovich 2017} considered a similar effect where stellar-mass BH binaries merging around a supermassive BH are embedded in a
non-spherical galactic potential.
They found that extreme eccentricity excitation is possible if the LK timescale driven by
the central massive BH is comparable to the nodal precession timescale of the binary centre of mass driven by the non-spherical potential.
An enhanced merger rate may also be achieved due to the effect of vector resonant relaxation of
BH binaries in galactic nuclei \citep[][]{Hamers 2018b}.

In this paper, we study binary BH mergers in quadruple systems (Figure \ref{fig:configuration}),
focusing on the initially wide ($\sim100$ AU) BH binaries as in \citet{Liu-ApJ}.
We show that binary-binary interactions increase the LK window for extreme eccentricity excitations, and
therefore significantly increase the BH binary merger fraction. We
quantify the parameter space (e.g., the orbital properties of the tertiary binary)
where this increase occurs.
Our result suggests that although the quadruple stellar systems
may not be as common as triples \citep[e.g.,][]{Sana},
they could be the dominant sources for dynamically enhanced BH mergers in the galactic field.

Our paper is organized as follows.
In Section \ref{sec 2}, we summarize the secular equations of motion in the octupole order
to evolve the quadruple systems with gravitational reaction. These equations are
based on the double-averaged approximation (averaging over both inner and outer orbits)
for the orbital evolution of hierarchical quadruple systems.
In Section \ref{sec 3}, we present the basic properties of LK oscillations
for general stellar quadruples.
In Section \ref{sec 4}, we perform a suite of numerical integrations to determine
the merger windows for LK-induced binary mergers,
assuming isotropic distribution of the orientations of tertiary binaries.
The associated merger fractions of BH binaries are then obtained.
We summarize our main results in Section \ref{sec 5}.

\section{Octupole-level Equations of Motion for Binary-Binary Systems}
\label{sec 2}

\begin{figure}
\centering
\begin{tabular}{cccc}
\includegraphics[width=8cm]{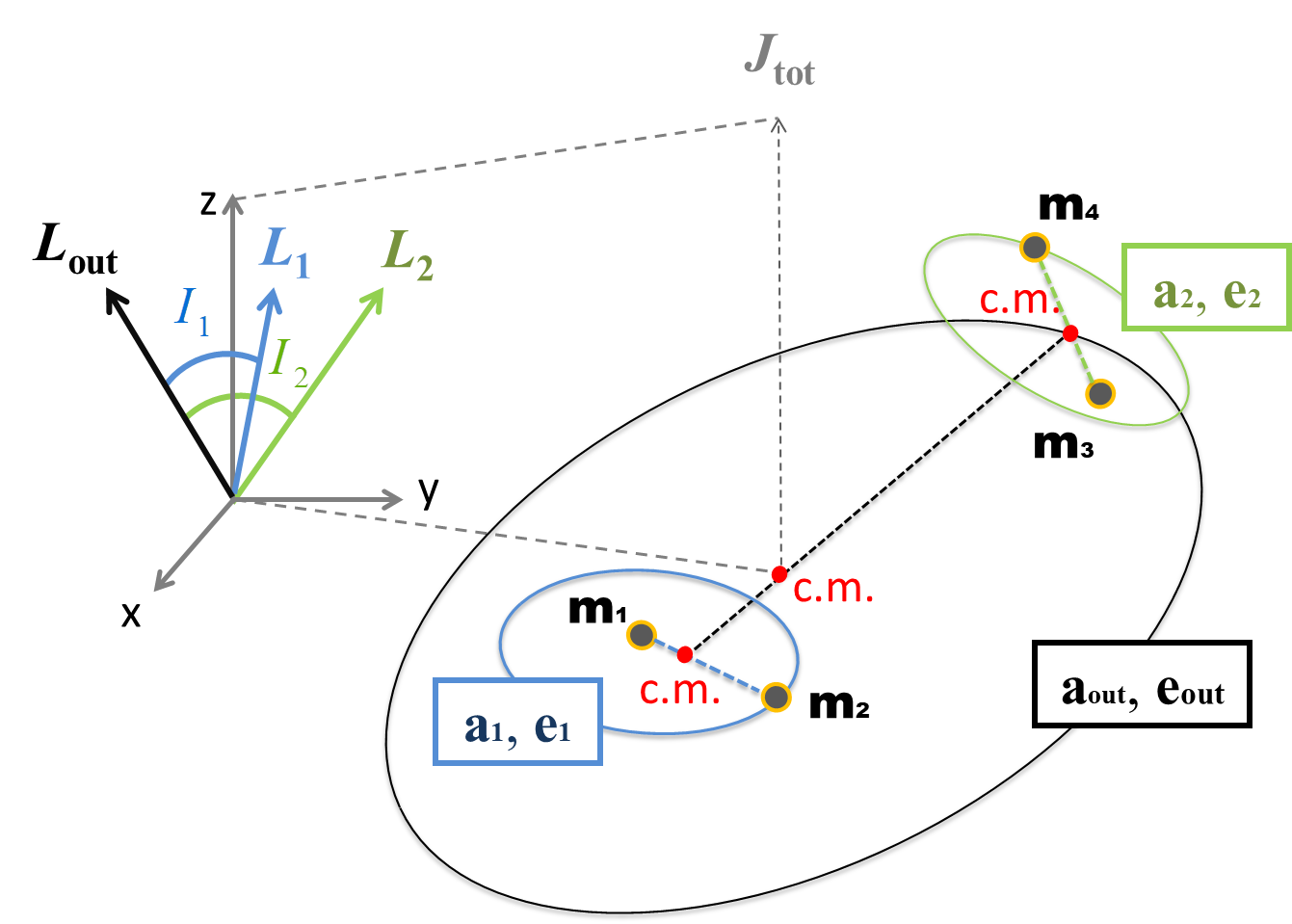}
\end{tabular}
\caption{Illustration of the binary-binary system.
The first (inner) binary is comprised of two BHs ($m_1$ and $m_2$);
the second binary consists of another two bodies ($m_3$ and $m_4$) and
orbits the center mass of the first inner binary, constituting the outer orbit.
Here, $a_{1,2,\OUT}$ are the semi-major axes, $e_{1,2,\OUT}$ are the eccentricities
of each binary. The total angular momentum $\textbf{J}_\tot=\textbf{L}_1+\textbf{L}_2+\textbf{L}_\OUT$
is along the $z$-axis, where $\textbf{L}_1$, $\textbf{L}_2$ and $\textbf{L}_\OUT$ (not to scale) denote
the angular momenta of the first, second (inner) binaries and outer orbit, respectively,
``c.m." indicates the center of mass of each system.
$I_1$ and $I_2$ are the mutual inclinations between $\textbf{L}_1$ and $\textbf{L}_\OUT$, $\textbf{L}_2$ and $\textbf{L}_\OUT$, respectively.
}
\label{fig:configuration}
\end{figure}

We consider a hierarchical quadruple system, composed of
two binaries orbiting each other, as depicted in Figure \ref{fig:configuration}.
The first (inner) BH binary has the masses $m_1$, $m_2$ and the distant second (inner) binary has the masses $m_3$ and $m_4$.
The reduced mass for the first binary is $\mu_1\equiv m_1m_2/m_{12}$, with $m_{12}\equiv m_1+m_2$
and the second binary has $\mu_2\equiv m_3m_4/m_{34}$, with $m_{34}\equiv m_3+m_4$.
The outer binary ($m_{12}$ orbits around $m_{34}$) has $\mu_\OUT\equiv(m_{12}m_{34})/m_\tot$ with $m_\tot\equiv m_{12}+m_{34}$.
The semi-major axes and eccentricities are denoted by $a_1$, $a_2$, $a_\OUT$ and $e_1$, $e_2$, $e_\OUT$, respectively.
The orbital angular momenta of three orbits are
\begin{eqnarray}
&&\textbf{L}_1=\mathrm{L}_1\hat{\textbf{L}}_1=\mu_1\sqrt{G m_{12}a_1(1-e_1^2)}\,\hat{\textbf{L}}_1,\\
&&\textbf{L}_2=\mathrm{L}_2\hat{\textbf{L}}_2=\mu_2\sqrt{G m_{34}a_2(1-e_2^2)}\,\hat{\textbf{L}}_2,\\
&&\textbf{L}_\OUT=\mathrm{L}_\OUT\hat{\textbf{L}}_\OUT=\mu_\OUT\sqrt{G m_\tot a_\OUT(1-e_\OUT^2)}\,\hat{\textbf{L}}_\OUT,
\end{eqnarray}
where $\hat{\bf L}_1$, $\hat{\bf L}_2$ and $\hat{\bf L}_\OUT$ are unit vectors.
We also define the eccentricity vectors as $\textbf{e}_1=e_1\hat{\textbf{e}}_1$,
$\textbf{e}_2=e_2\hat{\textbf{e}}_2$ and $\textbf{e}_\OUT=e_\OUT\hat{\textbf{e}}_\OUT$.
For simplicity, we only study the LK-induced orbital decay in the first inner binary,
considering the second one as an external perturber.
Thus, for convenience of notation, we will frequently omit the subscript ``$1$"
for the first inner binary.

The secular equations of motion for the two inner binaries take the form:
\begin{eqnarray}
&&\frac{d \textbf{L}}{dt}=\frac{d \textbf{L}}{dt}\bigg|_{\mathrm{LK}}+\frac{d \textbf{L}}{dt}\bigg|_{\mathrm{GW}}~,\label{eq:Full Kozai 1}\\
&&\frac{d \mathbf{e}}{dt}=\frac{d \mathbf{e}}{dt}\bigg|_{\mathrm{LK}}+\frac{d \mathbf{e}}{dt}\bigg|_{\mathrm{GR}}+\frac{d
\mathbf{e}}{dt}\bigg|_{\mathrm{GW}}~,\label{eq:Full Kozai 2}\\
&&\frac{d \textbf{L}_2}{dt}=\frac{d \textbf{L}_2}{dt}\bigg|_{\mathrm{LK}}~,\label{eq:Full Kozai 3}\\
&&\frac{d \mathbf{e}_2}{dt}=\frac{d \mathbf{e}_2}{dt}\bigg|_{\mathrm{LK}}+\frac{d \mathbf{e}_2}{dt}\bigg|_{\mathrm{GR}}~,\label{eq:Full Kozai 4}
\end{eqnarray}
and the outer orbit follows
\begin{eqnarray}
&&\frac{d \textbf{L}_\OUT}{dt}=\frac{d \textbf{L}_\OUT}{dt}\bigg|_\mathrm{1st}+\frac{d \textbf{L}_\OUT}{dt}\bigg|_\mathrm{2nd}\label{eq:jout},\\
&&\frac{d \textbf{e}_\OUT}{dt}=\frac{d \textbf{e}_\OUT}{dt}\bigg|_\mathrm{1st}+\frac{d \textbf{e}_\OUT}{dt}\bigg|_\mathrm{2nd}\label{eq:eout}.
\end{eqnarray}
In the first binary, we include the contributions from the outer binary (with perturber mass $m_{34}$)
that generate LK oscillations (subscripted by ``LK"),
the general relativistic (GR) post-Newtonian correction,
and the dissipation due to gravitational wave (GW) emission
(see Equations \ref{eq:Full Kozai 1}-\ref{eq:Full Kozai 2}).
We also evolve the second binary throughout the paper as the first one but without the GW radiation (Equations \ref{eq:Full Kozai 3}-\ref{eq:Full Kozai 4}).
The outer binary's angular momentum and eccentricity are affected by Newtonian potential from both the first and second inner binaries
(subscripted by ``1st" and ``2nd").

To describe the LK oscillations, we introduce the reduced angular momentum vectors as
\ba
&&\mathbf{j}\equiv j\hat{\textbf{L}}=\sqrt{1-e^2}\hat{\textbf{L}},\\
&&\mathbf{j}_\OUT\equiv j_\OUT\hat{\textbf{L}}_\OUT=\sqrt{1-e_\OUT^2}\hat{\textbf{L}}_\OUT.
\ea
Therefore, for the first binary, we have, to the octupole order \citep[][]{Liu et al 2015,Petrovich 2015}
\ba\label{eq:j1vec}
&&\frac{d{\jvec}}{dt}\bigg|_{\mathrm{LK}}=\frac{3}{4~t_{\lk,12}}\Big[(\jvec\cdot\nvec_\OUT)~\jvec\times\nvec_\OUT
-5(\evec\cdot\nvec_\OUT)~\evec\times\nvec_\OUT\Big]\nonumber\\
&&-\frac{75\varepsilon_{\oct,12}}{64~t_{\lk,12}}\Bigg\{
\bigg[2\Big[(\evec\cdot\uvec_\OUT)(\jvec\cdot\nvec_\OUT)\nonumber\\
&&+(\evec\cdot\nvec_\OUT)(\jvec\cdot\uvec_\OUT)\Big]~\jvec+2\Big[(\jvec\cdot\uvec_\OUT)(\jvec\cdot\nvec_\OUT)\nonumber\\
&&-7(\evec\cdot\uvec_\OUT)(\evec\cdot\nvec_\OUT)\Big]~\evec\bigg]\times\nvec_\OUT\nonumber\\
&&+\bigg[2(\evec\cdot\nvec_\OUT)(\jvec\cdot\nvec_\OUT)~\jvec
+\Big[\frac{8}{5}e^2-\frac{1}{5}\nonumber\\
&&-7(\evec\cdot\nvec_\OUT)^2+(\jvec\cdot\nvec_\OUT)^2\Big]~\evec\bigg]
\times\uvec_\OUT\Bigg\},
\ea
and
\ba\label{eq:e1vec}
&&\frac{d{\evec}}{dt}\bigg|_{\mathrm{LK}}=\frac{3}{4~t_{\lk,12}}\Big[(\jvec\cdot\nvec_\OUT)~\evec\times\nvec_\OUT
+2~\jvec\times\evec\nonumber\\
&&-5(\evec\cdot\nvec_\OUT)\jvec\times\nvec_\OUT\Big]\nonumber\\
&&-\frac{75\varepsilon_{\oct,12}}{64~t_{\lk,12}}\Bigg\{
\bigg[2(\evec\cdot\nvec_\OUT)(\jvec\cdot\nvec_\OUT)~\evec\nonumber\\
&&+\Big[\frac{8}{5}e^2-\frac{1}{5}-7(\evec\cdot\nvec_\OUT)^2+(\jvec\cdot\nvec_\OUT)^2\Big]~\jvec\bigg]\times\uvec_\OUT\nonumber\\
&&+\bigg[2\Big[(\evec\cdot\uvec_\OUT)(\jvec\cdot\nvec_\OUT)+(\evec\cdot\nvec_\OUT)(\jvec\cdot\uvec_\OUT)\Big]~\evec\nonumber\\
&&+2\Big[(\jvec\cdot\nvec_\OUT)(\jvec\cdot\uvec_\OUT)-7(\evec\cdot\nvec_\OUT)(\evec\cdot\uvec_\OUT)\Big]~\jvec
\bigg]\times\nvec_\OUT\nonumber\\
&&+\frac{16}{5}(\evec\cdot\uvec_\OUT)~\jvec\times\evec\Bigg\}~,
\ea
where
\be\label{eq:varepsilon oct}
\varepsilon_{\rm oct,12}\equiv {m_1-m_2\over m_{12}}\left({a\over a_\OUT}\right)
{e_\OUT\over 1-e_\OUT^2}
\ee
measures the relative strength of the octupole potential compared to the quadrupole one.
The quadrupole term induces the oscillations in the eccentricity
and mutual orbital inclination on the timescale of
\be\label{eq:LK timescale}
t_{\lk,12}=\frac{1}{n}\frac{m_{12}}{m_{34}}\bigg(\frac{a_{\OUT,\eff}}{a}\bigg)^3,
\ee
where $n=(G m_{12}/a^3)^{1/2}$ is the mean motion of the first inner binary and
the effective outer binary separation is defined as
\be\label{eq:aout eff}
a_{\OUT,\eff}\equiv a_\OUT\sqrt{1-e^2_\OUT}.
\ee

General Relativity (1-PN correction) introduces pericenter precession as
\be\label{eq:e GR}
\frac{d \mathbf{e}}{dt}\bigg|_{\mathrm{GR}}=\Omega_\mathrm{GR}\hat{\textbf{L}}\times\mathbf{e},
\ee
with the precession rate given by
\be\label{eq:GR}
\Omega_\mathrm{GR}=\frac{3Gn m_{12}}{c^2a(1-e^2)},
\ee

Gravitational radiation draws energy and angular momentum from the BH orbit .
The rates of change of $\textbf{L}$ and $\mathbf{e}$ are \citep[]{Peters 1964}
\begin{eqnarray}
&&\frac{d \textbf{L}}{dt}\bigg|_{\mathrm{GW}}=-\frac{32}{5}\frac{G^{7/2}}{c^5}\frac{\mu^2 m_{12}^{5/2}}{a^{7/2}}
\frac{1+7e^2/8}{(1-e^2)^2}\hat{\textbf{L}},\label{eq:GW 1}\\
&&\frac{d \mathbf{e}}{dt}\bigg|_{\mathrm{GW}}=-\frac{304}{15}\frac{G^3}{c^5}\frac{\mu m_{12}^2}{a^4(1-e^2)^{5/2}}
\bigg(1+\frac{121}{304}e^2\bigg)\mathbf{e}.\label{eq:GW 2}
\end{eqnarray}
The merger time due to GW radiation of an isolated binary with the
initial semi-major axis $a_0$ and eccentricity $e_0=0$ is given by
\ba\label{eq:Tmerger}
&&T_\mathrm{m,0}=\frac{5c^5 a_0^4}{256 G^3 m_{12}^2 \mu}\\
&&~~~~~~\simeq10^{10}\bigg(\frac{60M_\odot}{m_{12}}\bigg)^2
\bigg(\frac{15M_\odot}{\mu}\bigg)\bigg(\frac{a_0}{0.202\au}\bigg)^4\mathrm{yrs}\nonumber.
\ea

In our calculations, we also evolve the second binary, except that we do not include the GW terms for the sake of clarity.
By switching the indices $\mathbf{j}\rightarrow\mathbf{j}_2$, $\mathbf{e}\rightarrow\mathbf{e}_2$,
$\varepsilon_{\rm oct,34}$ and $t_{\lk,34}$,
Equations (\ref{eq:j1vec})-(\ref{eq:e1vec}) an (\ref{eq:e GR}) can be applied to the second binary
(with
$m_1\rightarrow m_3$, $m_2\rightarrow m_4$, $m_{12}\rightarrow m_{34}$,
$a\rightarrow a_2$ and $n\rightarrow(G m_{34}/a_2^3)^{1/2}$ in Equations \ref{eq:varepsilon oct}-\ref{eq:LK timescale}, \ref{eq:GR}).
The outer orbit is influenced by both first and second binary.
The first piece of Equation (\ref{eq:jout}) is given by
\ba\label{eq:j2vec 1st}
&&\frac{d{\jvec_\OUT}}{dt}\bigg|_\mathrm{1st}=\frac{3}{4t_{\lk,12}}
\frac{\Lambda}{\Lambda_\OUT}\Big[(\jvec\cdot\nvec_\OUT)~\nvec_\OUT\times\jvec\nonumber\\
&&-5(\evec\cdot\nvec_\OUT)~\nvec_\OUT\times\evec\Big]\nonumber\\
&&-\frac{75\varepsilon_{\oct,12}}{64t_{\lk,12}}\frac{\Lambda}{\Lambda_\OUT}\Bigg\{
2\Big[(\evec\cdot\nvec_\OUT)(\jvec\cdot\uvec_\OUT)~\nvec_\OUT\nonumber\\
&&+(\evec\cdot\uvec_\OUT)(\jvec\cdot\nvec_\OUT)~\nvec_\OUT
+(\evec\cdot\nvec_\OUT)(\jvec\cdot\nvec_\OUT)~\uvec_\OUT\Big]
\times\jvec\nonumber\\
&&+\bigg[2(\jvec\cdot\uvec_\OUT)(\jvec\cdot\nvec_\OUT)~\nvec_\OUT
-14(\evec\cdot\uvec_\OUT)(\evec\cdot\nvec_\OUT)~\nvec_\OUT\nonumber\\
&&+\Big[\frac{8}{5}e^2-\frac{1}{5}
-7(\evec\cdot\nvec_\OUT)^2+(\jvec\cdot\nvec_\OUT)^2\Big]~\uvec_\OUT\frac{}{}\bigg]\times\evec
\Bigg\}.
\ea
The evolution equation of $\mathbf{L}_\OUT$ is
$(d \textbf{L}_\OUT/dt)|_{\mathrm{1st}}=\mu_\OUT\sqrt{Gm_\tot a_\OUT}~(d \textbf{j}_\OUT/dt)|_{\mathrm{1st}}$. Also
\ba\label{eq:e2vec 1st}
&&\frac{d{\evec_\OUT}}{dt}\bigg|_\mathrm{1st}=\frac{3}{4t_{\lk,12}\sqrt{1-e_\OUT^2}}\frac{\Lambda}{\Lambda_\OUT}\bigg[
(\jvec\cdot\nvec_\OUT)~\evec_\OUT\times\jvec\nonumber\\
&&-5(\evec\cdot\nvec_\OUT)\evec_\OUT\times\evec-\Big[\frac{1}{2}-3e^2+\frac{25}{2}(\evec\cdot\nvec_\OUT)^2\nonumber\\
&&-\frac{5}{2}(\jvec\cdot\nvec_\OUT)^2\Big]\nvec_\OUT\times\evec_\OUT\bigg]
-\frac{75}{64t_{\lk,12}}\frac{\varepsilon_{\oct,12}}{\sqrt{1-e_\OUT^2}}\frac{\Lambda}{\Lambda_\OUT}\nonumber\\
&&\times\Bigg\{2\Big[(\evec\cdot\nvec_\OUT)(\jvec\cdot\evec_\OUT)~\uvec_\OUT
+(\jvec\cdot\nvec_\OUT)(\evec\cdot\evec_\OUT)~\uvec_\OUT\nonumber\\
&&+\frac{1-e_\OUT^2}{e_\OUT}(\evec\cdot\nvec_\OUT)
(\jvec\cdot\nvec_\OUT)~\nvec_\OUT\Big]
\times\jvec\nonumber\\
&&+\bigg[2(\jvec\cdot\evec_\OUT)(\jvec\cdot\nvec_\OUT)~\uvec_\OUT
-14(\evec\cdot\evec_\OUT)(\evec\cdot\nvec_\OUT)~\uvec_\OUT\nonumber\\
&&+\frac{1-e_\OUT^2}{e_\OUT}\Big[\frac{8}{5}e^2
-\frac{1}{5}-7(\evec\cdot\nvec_\OUT)^2\nonumber\\
&&+(\jvec\cdot\nvec_\OUT)^2\Big]~\nvec_\OUT\bigg]\times\evec-\bigg[2\left(\frac{1}{5}-\frac{8}{5}e^2\right)(\evec\cdot\uvec_\OUT)~\evec_\OUT\nonumber\\
&&+14(\evec\cdot\nvec_\OUT)(\jvec\cdot\uvec_\OUT)(\jvec\cdot\nvec_\OUT)~\evec_\OUT\nonumber\\
&&+7(\evec\cdot\uvec_\OUT)\Big[\frac{8}{5}e^2
-\frac{1}{5}-7(\evec\cdot\nvec_\OUT)^2\nonumber\\
&&+(\jvec\cdot\nvec_\OUT)^2\Big]~\evec_\OUT
\bigg]\times\nvec_\OUT\Bigg\}~~.
\ea
Here, we have defined
\ba
&&\Lambda\equiv \mathrm{L}|_{e=0}=\mu\sqrt{Gm_{12}a},\\
&&\Lambda_\OUT\equiv \mathrm{L}_\OUT|_{e_\OUT=0}=\mu_\OUT\sqrt{Gm_\tot a_\OUT}.
\ea
Similar expressions apply to $(d\jvec_\OUT/dt)|_\mathrm{2nd}$ and $(d\evec_\OUT/dt)|_\mathrm{2nd}$.
Equations (\ref{eq:Full Kozai 1})-(\ref{eq:eout}) completely determine the
secular evolution of the binary-binary system. These equations are based on the double averaging approximation, and
require that the timescale near the maximum eccentricity $e_\m$ be longer than the period of the outer binary
\citep[e.g.,][]{Seto PRL,Antonini 2014}, i.e.
\be\label{eq:DA condition}
t_\lk\sqrt{1-e_\m^2}\gtrsim P_\OUT.
\ee
See \citet{Liu-ApJ} for more discussion on the regime of validity of the
double-averaged equations and the more general single-averaged equations.

\section{Excitation of Eccentricity in Binary-Binary Systems}
\label{sec 3}

\begin{figure*}
\centering
\begin{tabular}{cc}
\includegraphics[width=8cm]{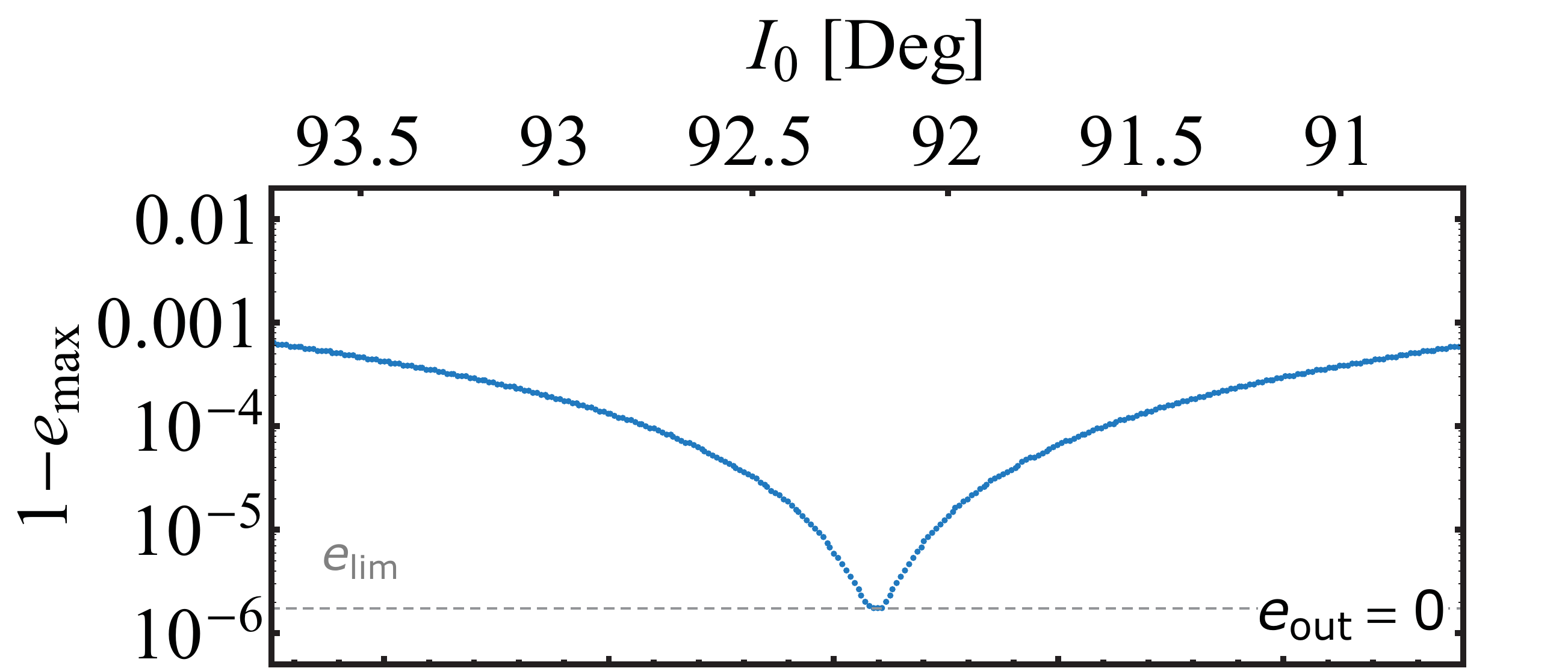}&
\includegraphics[width=8cm]{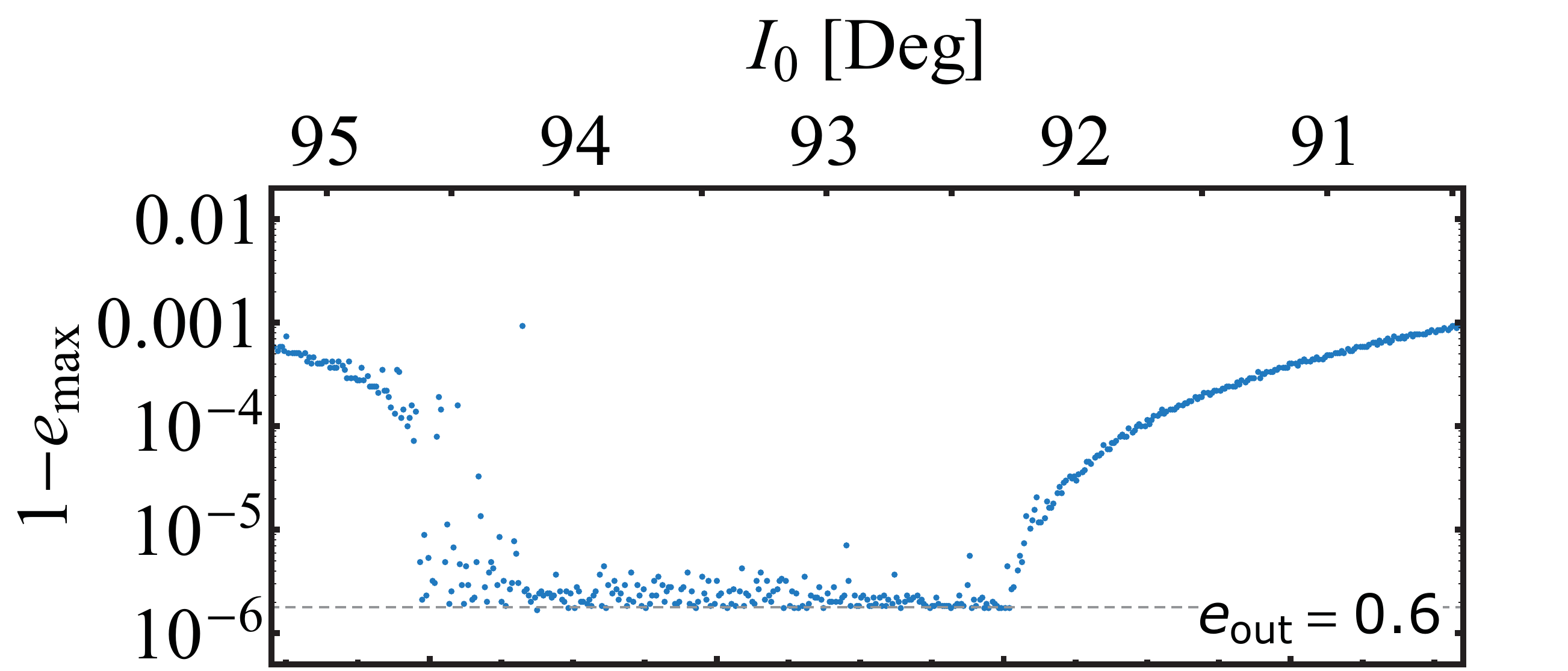}\\
\includegraphics[width=8cm]{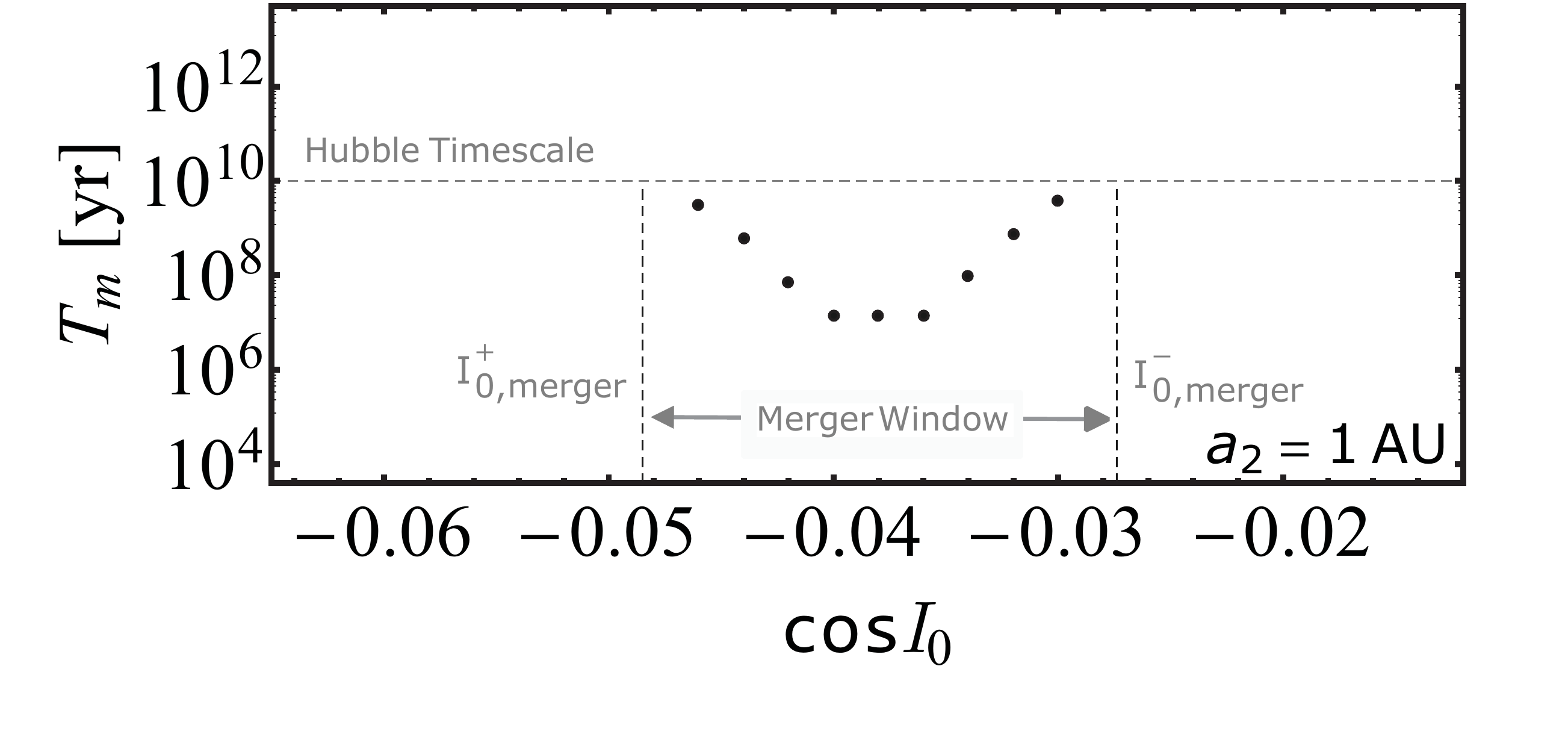}&
\includegraphics[width=8cm]{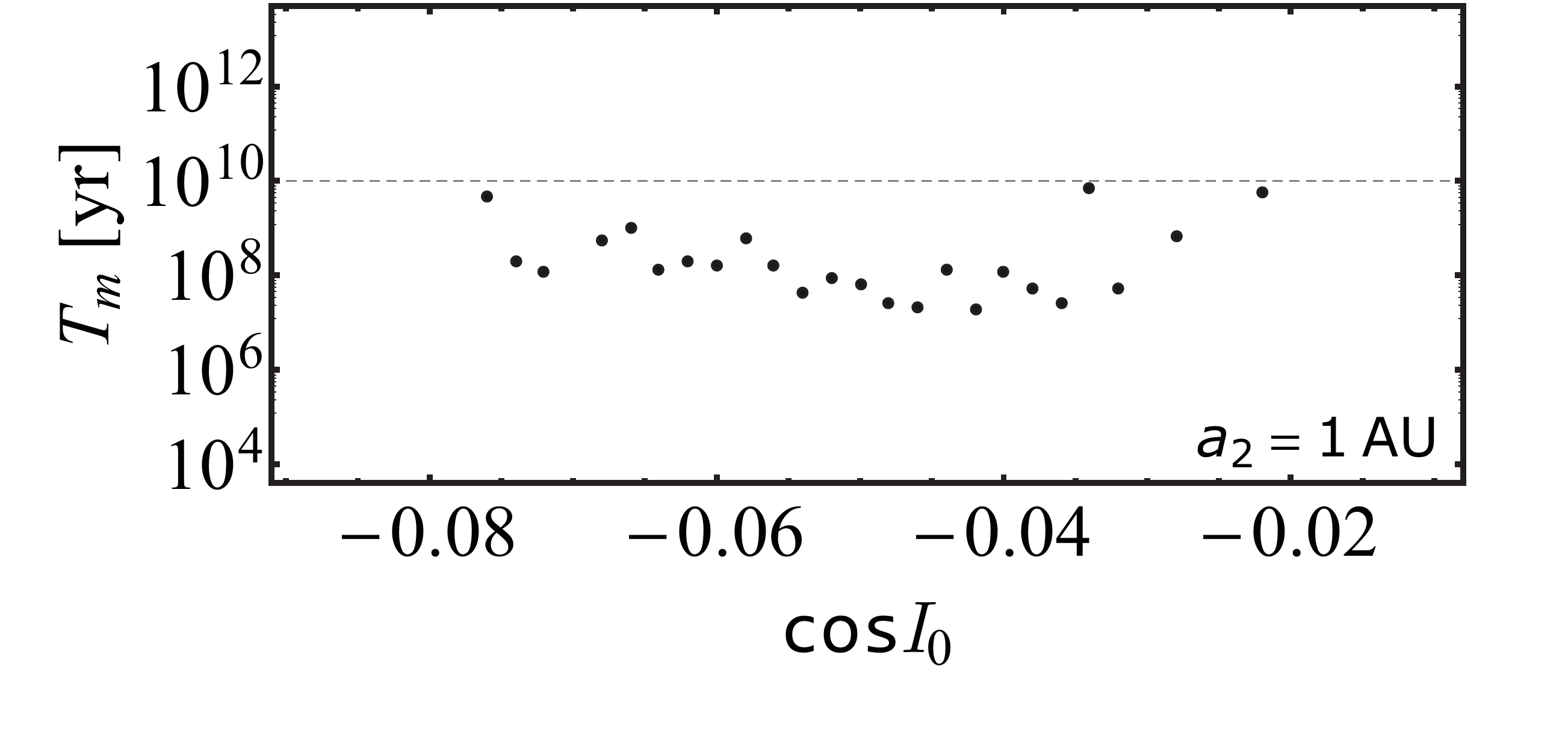}\\
\\
\\
\includegraphics[width=8cm]{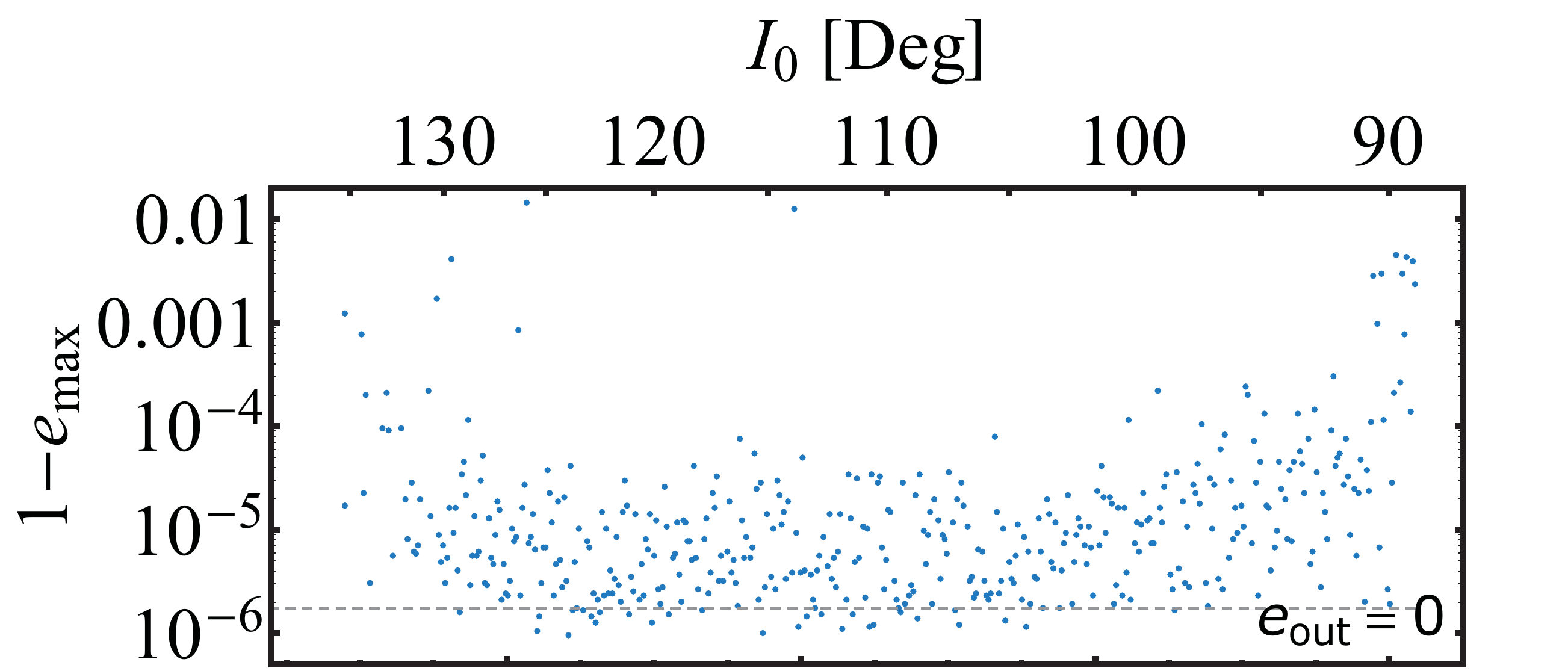}&
\includegraphics[width=8cm]{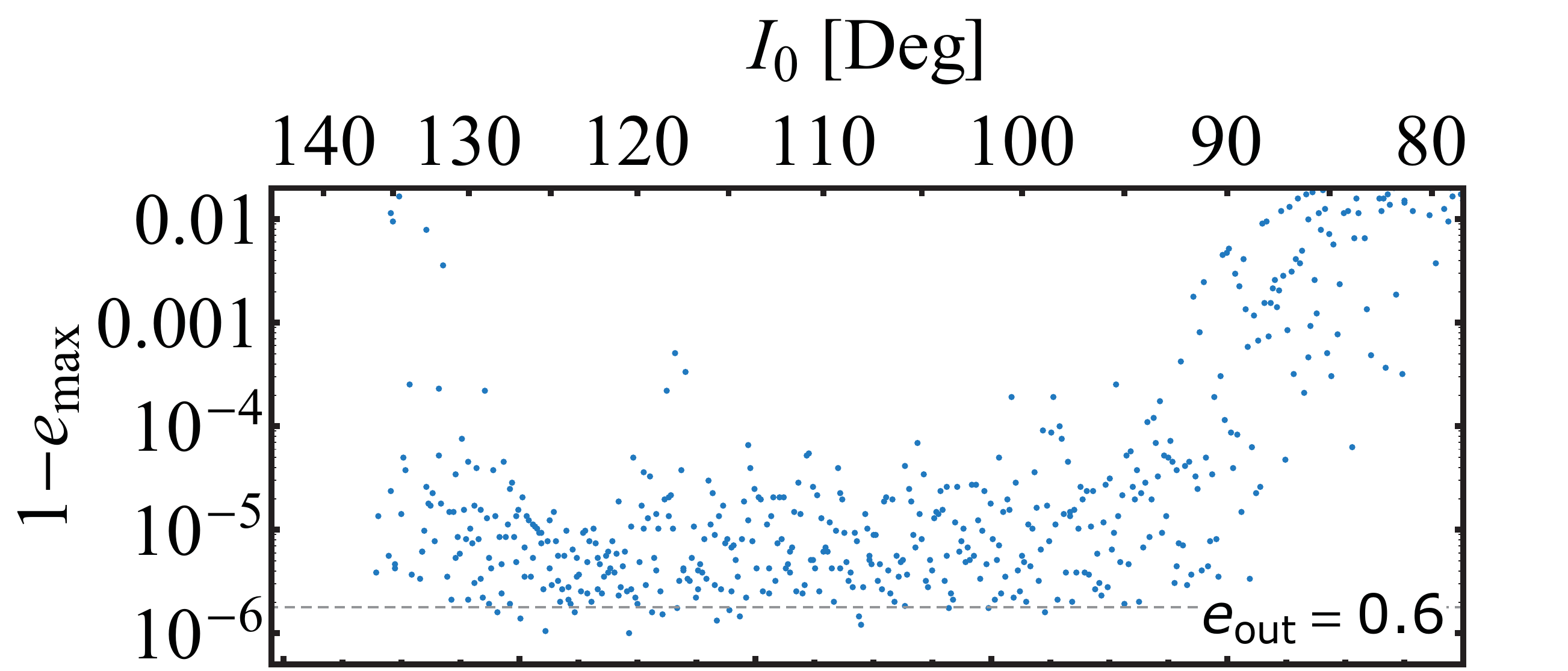}\\
\includegraphics[width=8cm]{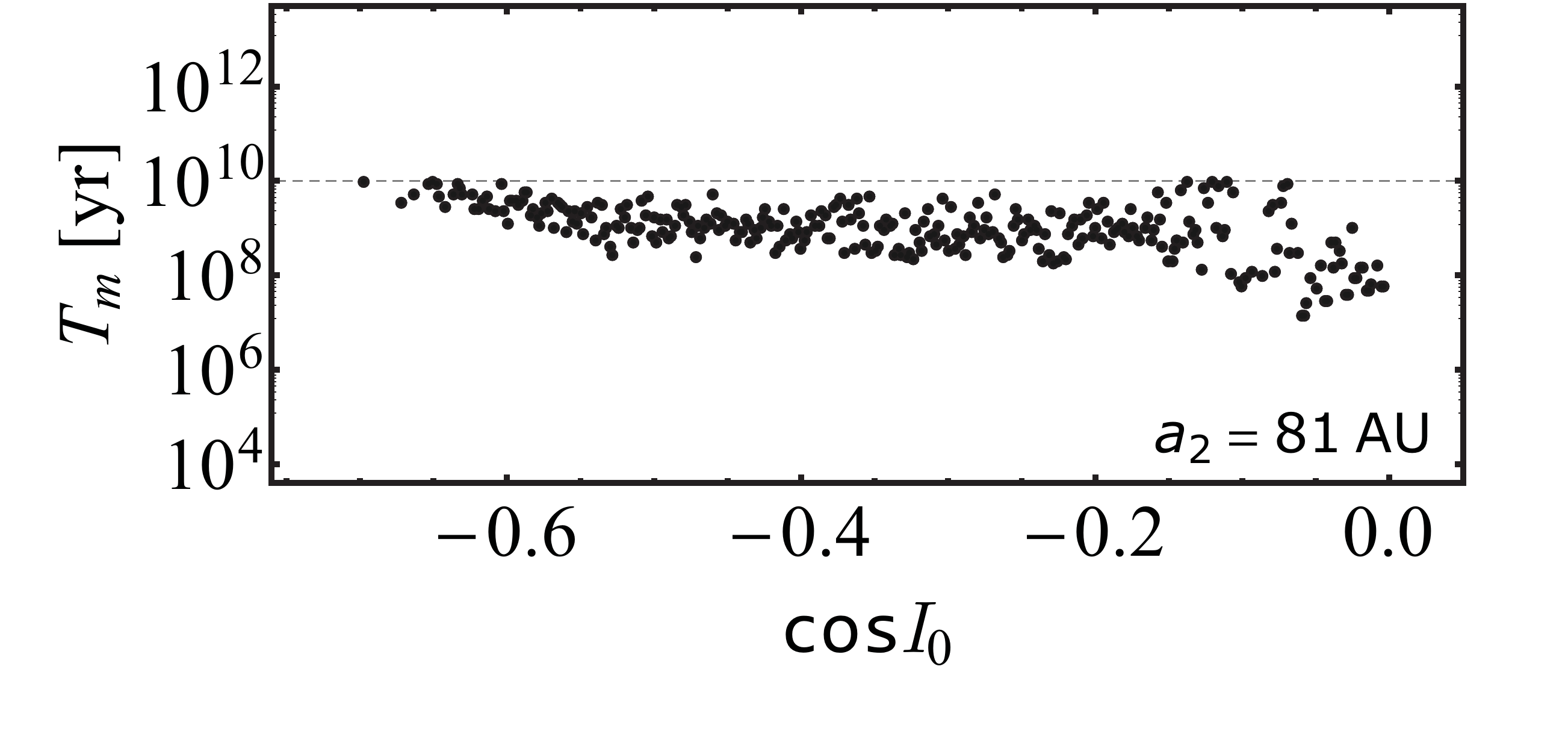}&
\includegraphics[width=8cm]{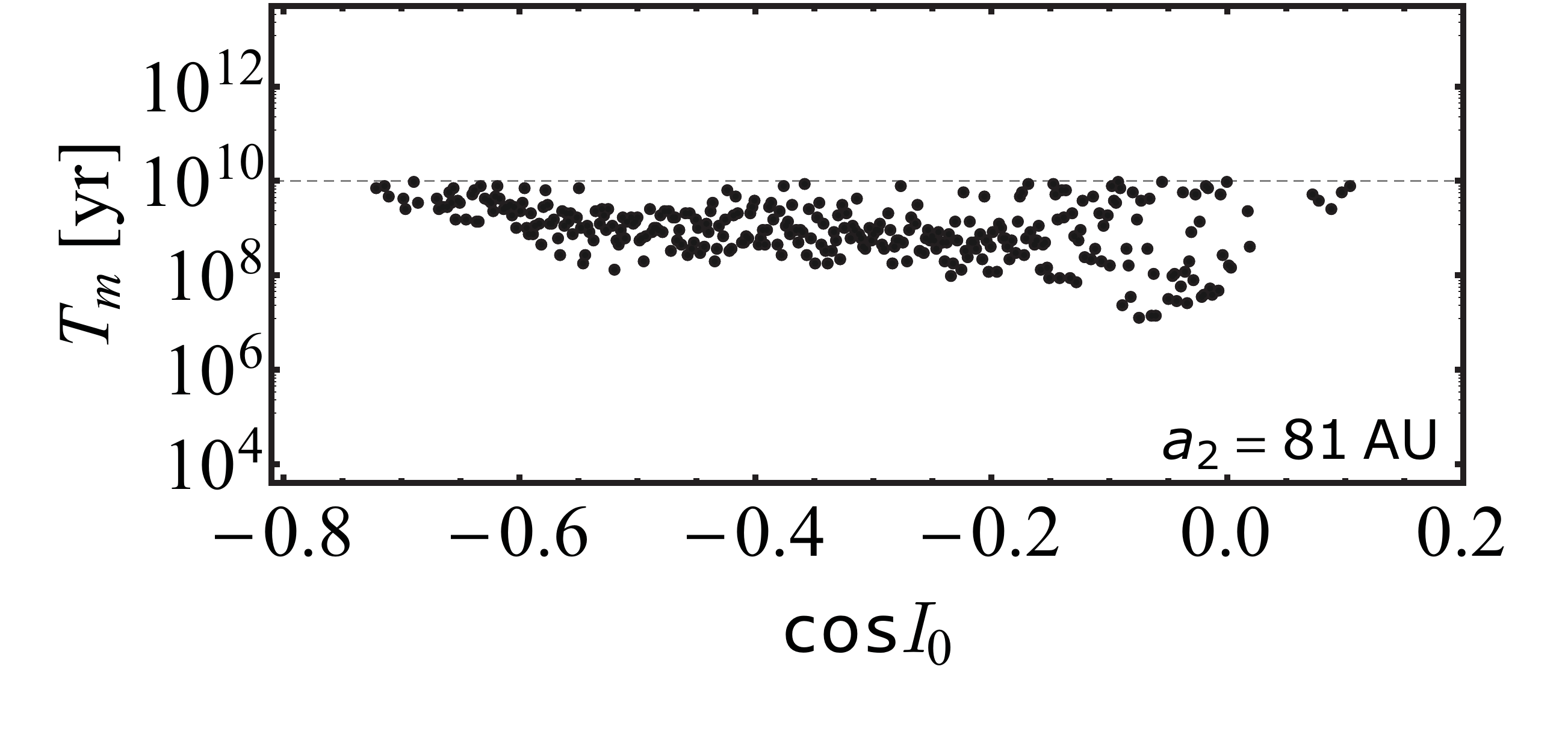}\\
\end{tabular}
\caption{Eccentricity excitation and merger window in binary-binary systems for different values of $e_\OUT$ and $a_2$.
In each panel, the upper and lower plots show the maximum eccentricity $e_\m$ (assuming no GW emission),
and the first (inner) binary merger time $T_\mathrm{m}$ (with GW emission)
as a function of $I_0$ (the initial value of $I_1$).
The system parameters are: $m_1=30M_\odot$, $m_2=20M_\odot$, $a_0=100\au$
(initial value of $a$), $e_0=e_{2,0}=0.001$, $m_3=m_4=15M_\odot$, $a_\OUT=4400\au$ (for $e_\OUT=0$) and $a_\OUT=5500\au$ (for $e_\OUT=0.6$).
The longitude of the periapse $\omega_{\OUT,0}$ is randomly chosen in the range of $(0, 2\pi)$.
All four panels have the same $\bar{a}_{\OUT,\eff}=4.4$ (Equation \ref{eq:aout bar}), implying the same
quadrupolar perturbation strength from the second tertiary binary, and the double-averaged secular approximation can be safely satisfied
(Equation \ref{eq:DA condition}).
Each black dot (obtained by solving Equations \ref{eq:Full Kozai 1}-\ref{eq:eout})
represents a successful merger event within $10^{10}$ yrs
(note that the grid of blue dots is denser than the black dots).
The dashed horizontal line ($e_\li$) is given by Equation (\ref{eq:ELIM}).
}
\label{fig:merger window}
\end{figure*}

Before considering the population of binary mergers in quadruple systems
(Section \ref{sec 4}), we first examine how binary-binary interaction influences the excitation of eccentricity
in the inner binary.

Figure \ref{fig:merger window} shows the maximum excited eccentricity achieved in the first binary ($e_\m$;
in the absence of GW emission) and merger window (including GW emission; to be discussed in Section \ref{sec 4})
as a function of the initial mutual inclination angle $I_0$ (the initial value of $I_1$).
The binary has masses $m_1=30M_\odot$, $m_2=20M_\odot$, and the initial semimajor axis $a_0=100\au$ and
the initial eccentricity $e_0=0$.
As noted before (see footnote \ref{fn:1}), such binaries cannot merge by themselves with tertiary companions.
For simplicity, we choose the second binary to have equal masses, i.e., $m_3=m_4=15M_\odot$
and circular orbit at $t=0$ ($e_{2,0}=0$).
Note that the unequal masses in the second binary can induce the similar behaviors as the first binary,
where the octupole contribution comes into play.
In this situation, both the evolution of $\hat{\mathbf{L}}_2$ and $\hat{\mathbf{L}}_\OUT$ might become chaotic,
which in turn could widen the window of $e$-excitation in the first inner binary.
The properties of the dynamics is beyond the scope of this paper and we leave it to a future work.
Here, we integrate the Equations (\ref{eq:Full Kozai 1})-(\ref{eq:eout}) and fix the initial inclination of the second binary to be $I_{2,0}=30^\circ$,
so that no LK oscillations occur in the second binary,
and we concentrate on the eccentricity excitation of the first binary.
As in \citet{Liu-ApJ}, we introduce the effective outer binary semimajor axis
as $a_{\OUT,\eff}=a_\OUT\sqrt{1-e_\OUT^2}$ and define
\be\label{eq:aout bar}
\begin{split}
\bar{a}_{\OUT,\eff}&\equiv \bigg(\frac{a_{\OUT,\eff}}{1000\au}\bigg)\bigg(\frac{m_{34}}{30M_\odot}\bigg)^{-1/3}\\
&=\bigg(\frac{a_\OUT\sqrt{1-e_\OUT^2}}{1000\au}\bigg)\bigg(\frac{m_{34}}{30M_\odot}\bigg)^{-1/3}.
\end{split}
\ee
This quantity characterizes the ``quadrupole strength" of the outer perturber $m_{34}=m_3+m_4$.
In the examples depicted in Figure \ref{fig:merger window}, we adopt
$\bar{a}_{\OUT,\eff}=4.4$, where the double-averaged secular equations
derived here can be safely used based on Equation (\ref{eq:DA condition}) (where we replace $e_\m$ with $e_\li$; see Equation \ref{eq:ELIM}).
Thus, we have $a_\OUT=4400\au$ for $e_\OUT=0$ and $a_\OUT=5500\au$ for $e_\OUT=0.6$.
The initial longitude of the periapse $\omega_\OUT$ is randomly chosen in the range of $(0, 2\pi)$.

The top two panels of Figure \ref{fig:merger window} show the results when $a_2\ll a_\OUT$.
In these cases, the binary-bianry system effectively reduces to a triple system,
with the first binary perturbed by $m_{34}$.
When $e_\OUT=0$ (the top left panel of Figure \ref{fig:merger window}), the octupole effect vanishes,
and the maximum eccentricity $e_\m$ achieved by the first binary (starting from $e_0\simeq0$)
can be evaluated analytically \citep[]{Liu et al 2015,Anderson et al 2017}:
\ba
&&\!\!\!\frac{3}{8}\frac{j^2_\mi-1}{j^2_\mi}\bigg[5\left(\cos I_0+\frac{\eta}{2}\right)^2-
\Bigl(3+4\eta\cos I_0+\frac{9}{4}\eta^2\Bigr)j^2_\mi \nonumber\\
&&\quad +\eta^2j^4_\mi\bigg]+\varepsilon_\gr \left(1-j_\mi^{-1}\right)=0,
\label{eq:EMAX}
\ea
where $j_\mi\equiv\sqrt{1-e_\m^2}$, $\eta\equiv(L/L_\OUT)_{e=0}$, and
\ba\label{eq:epsilonGR}
&&\varepsilon_\gr= \frac{3Gm_{12}^2a_{\OUT,\eff}^3}{c^2a^4m_{34}}\\
&&\simeq3.6\times10^{-5}\bigg(\!\frac{m_{12}}{60M_\odot}\!\bigg)^{\!\!2}\!\bigg(\!\frac{m_{34}}{30M_\odot}\!\bigg)^{\!\!-1}
\!\bigg(\!\frac{a_{\OUT,\eff}}{10^3\au}\!\bigg)^{\!\!3}\!\bigg(\!\frac{a}{10^2\au}\!\bigg)^{\!\!-4}\nonumber,
\ea
which measures the strength of the GR precession
(relative to the LK oscillations).
Note that in the limit of $\eta\rightarrow0$ and $\varepsilon_\gr\rightarrow0$,
Equation (\ref{eq:EMAX}) yields the well-known relation $e_\m=\sqrt{1-(5/3)\cos^2 I_0}$.
The maximum possible $e_\m$ for all values of $I_0$, called $e_\li$, is given by
\be\label{eq:ELIM}
\frac{3}{8}(j_\li^2-1)\left[-3+\frac{\eta^2}{4}\left(\frac{4}{5}j_\li^2-1\right)\right]+
\varepsilon_\gr \left(1-j_\li^{-1}\right)=0.
\ee
From the top panels of Figure \ref{fig:merger window} (with $a_2=1$ AU), we see that for $e_\OUT=0$,
the limiting eccentricity can be achieved only in a very narrow inclination window around $I_0=92.2^\circ$.
For $e_\OUT=0.6$ (corresponding to $\varepsilon_{\oct,12}\simeq0.003$),
the same limiting eccentricity applies \citep[see][]{Liu et al 2015}, but it can be achieved over a wide range of $I_0\in[92^\circ, 94.5^\circ]$.

The lower panels of Figure \ref{fig:merger window} show $e_\m$ versus $I_0$ when the second binary has a semimajor
axis $a_2=81\au$. We see that regardless of the value of $e_\OUT$ (i.e., the strength of the octupole potential),
extreme eccentricity excitation can be achieved over a much wider range of inclinations,
roughly from $90^\circ$ to $130^\circ$.

The enhanced inclination range for LK oscillations in binary-binary systems can be understood
as a resonance phenomenon \citep[][]{Hamers and Lai 2017}. Considering the simple
case where the second binary does not experience LK oscillation and stays circular ($e_2=0$) and
the outer binary is also circular ($e_\OUT=0$).
So, no octupole effect comes into play,
and the angular momentum axis of the outer binary is affected by the second binary via
\be\label{eq:j2vec 2nd}
\frac{d{\hat{\mathbf{L}}_\OUT}}{dt}\bigg|_\mathrm{2nd}=\frac{3}{4t_{\lk,34}}
\frac{L_2}{L_\OUT}\big(\hat{\mathbf{L}}_2\cdot\nvec_\OUT\big)~\nvec_\OUT\times\hat{\mathbf{L}}_2~~,
\ee
where $t_{\lk,34}$ is the LK timescale in the second binary,
given by
\be
t_{\lk,34}=\frac{1}{n_2}\frac{m_{34}}{m_{12}}\bigg(\frac{a_{\OUT,\eff}}{a_2}\bigg)^3,
\ee
where $n_2=(G m_{34}/a_2^3)^{1/2}$.
Thus, $\hat{\mathbf{L}}_\OUT$ is driven into precession around the
$\mathbf{L}_{2+\OUT}\equiv\mathbf{L}_2+\mathbf{L}_\OUT$ axis at the rate
\be
\Omega_\OUT=\frac{3}{4 t_{\lk,34}}\frac{|\mathbf{L}_2+\mathbf{L}_\OUT|}{L_\OUT} \big(\hat{\mathbf{L}}_2\cdot\nvec_\OUT\big)
\simeq\frac{3}{4 t_{\lk,34}}\cos I_2.
\ee
On the other hand, the outer binary drives LK oscillations of the (first) inner binary on timescale $t_{\lk,12}$.
Thus, we define the dimensionless parameter
\be\label{eq:beta}
\beta\equiv\Omega_\OUT t_{\lk,12}=\frac{3}{4}\cos I_2\bigg(\frac{a_2}{a_1}\bigg)^{3/2}
\bigg(\frac{m_1+m_2}{m_3+m_4}\bigg)^{3/2}.
\ee
The value of $\beta$ measures the ratio between the LK timescale in the first binary and the
precession timescale of the outer orbit.
When $\beta\ll1$, the second binary essentially acts like a single mass ($m_3+m_4$),
and ``normal" LK oscillations apply.
When $\beta\gg1$, $\hat{\mathbf{L}}_\OUT$ precesses rapidly around the $\mathbf{L}_{2+\OUT}$,
the problem again reduces to that of ``normal" LK oscillations, with $\hat{\mathbf{L}}_{2+\OUT}$ serving
as the effective $\hat{\mathbf{L}}_\OUT$. When $\beta\sim 1$, a secular resonance occurs that generates
large $I$ even for initially low-inclination systems, and this resonantly excited
inclination then leads to LK oscillations of the inner binary.

\begin{figure*}
\centering
\begin{tabular}{cc}
\includegraphics[width=8cm]{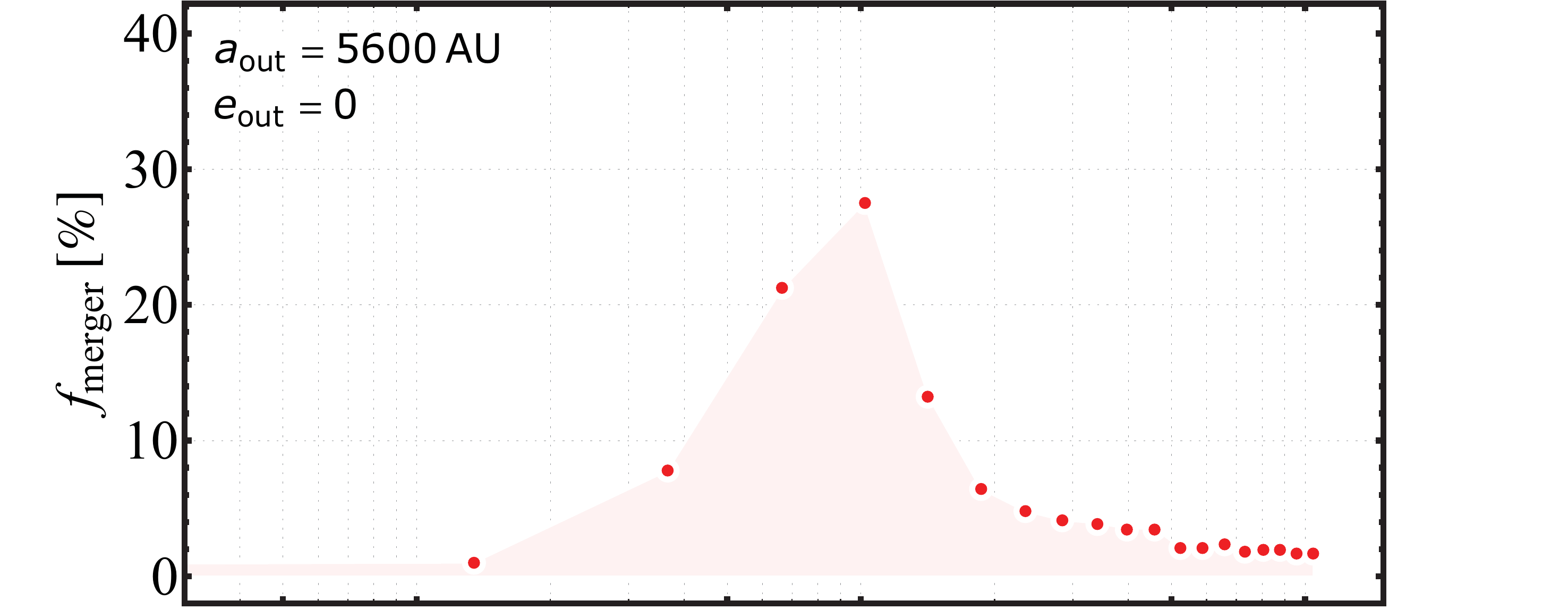}&
\includegraphics[width=8cm]{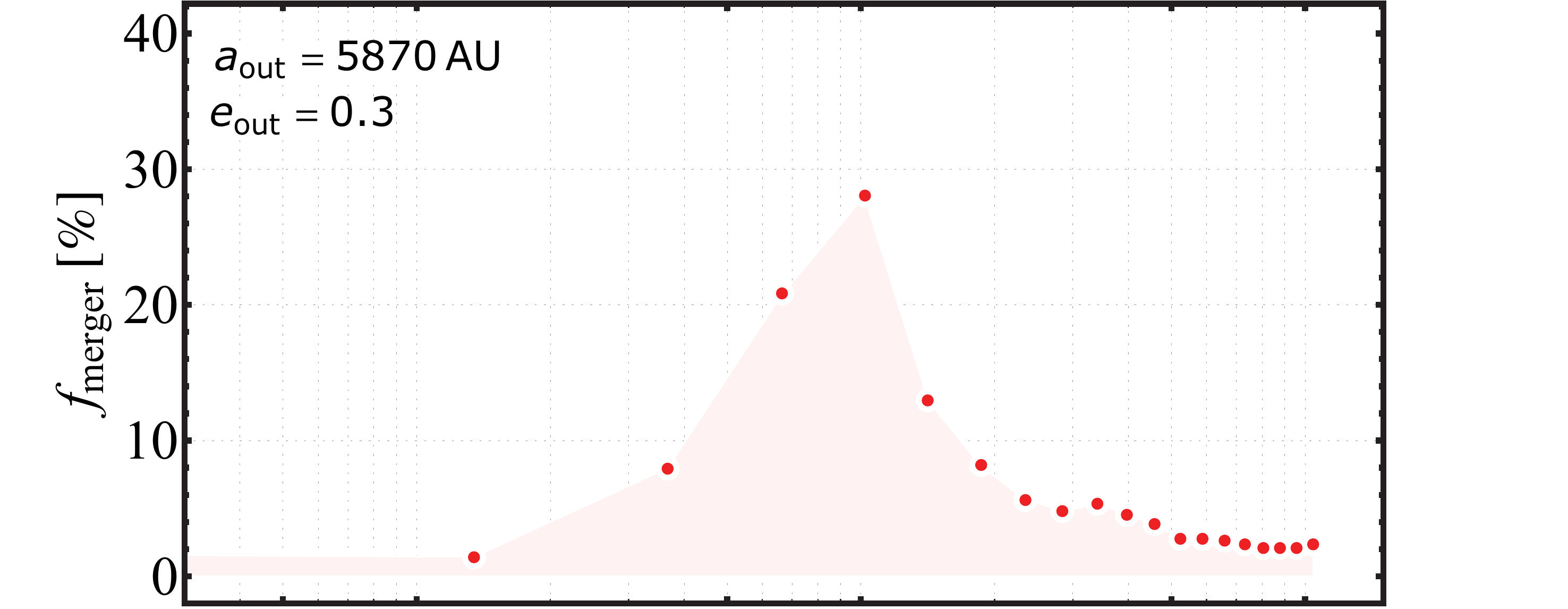}\\
\includegraphics[width=8cm]{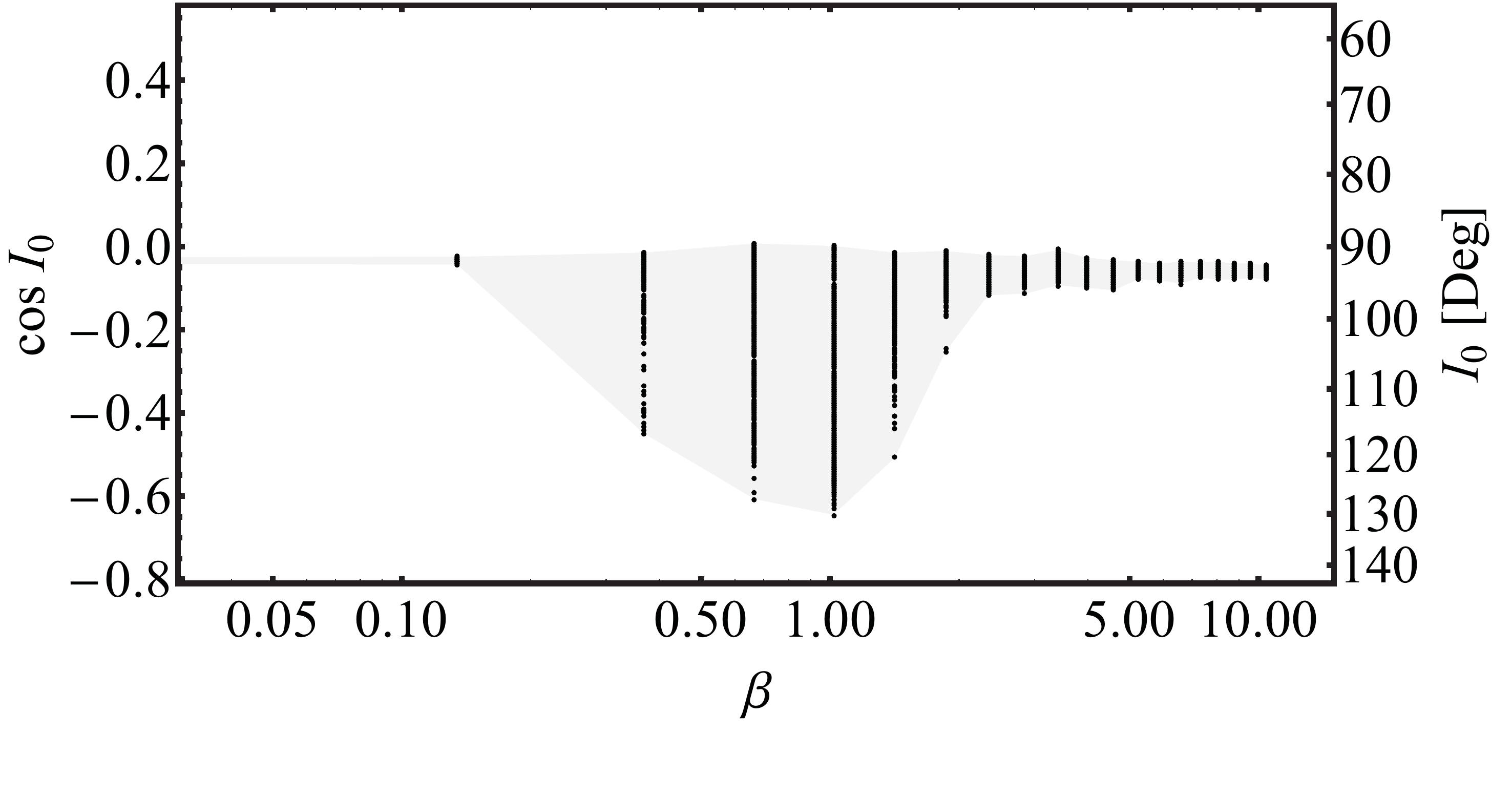}&
\includegraphics[width=8cm]{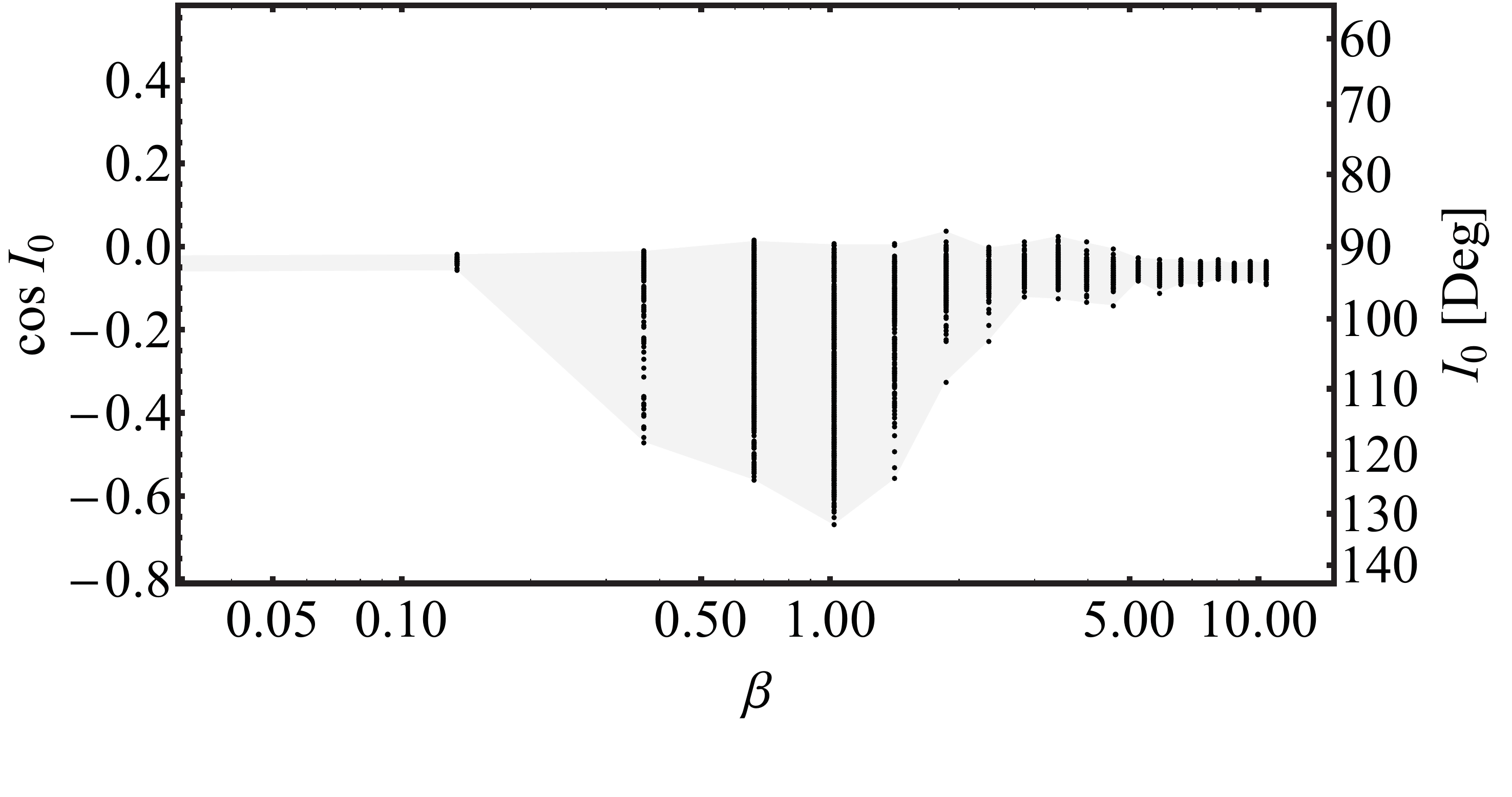}\\
\includegraphics[width=8cm]{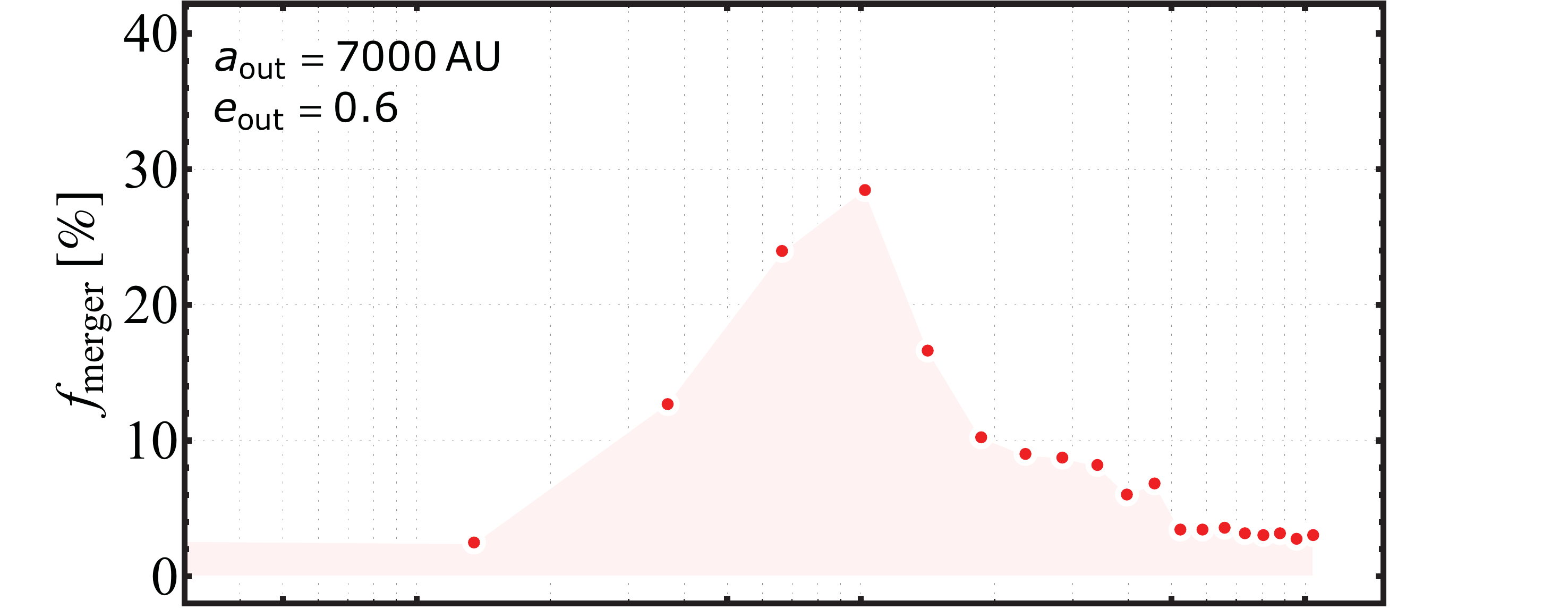}&
\includegraphics[width=8cm]{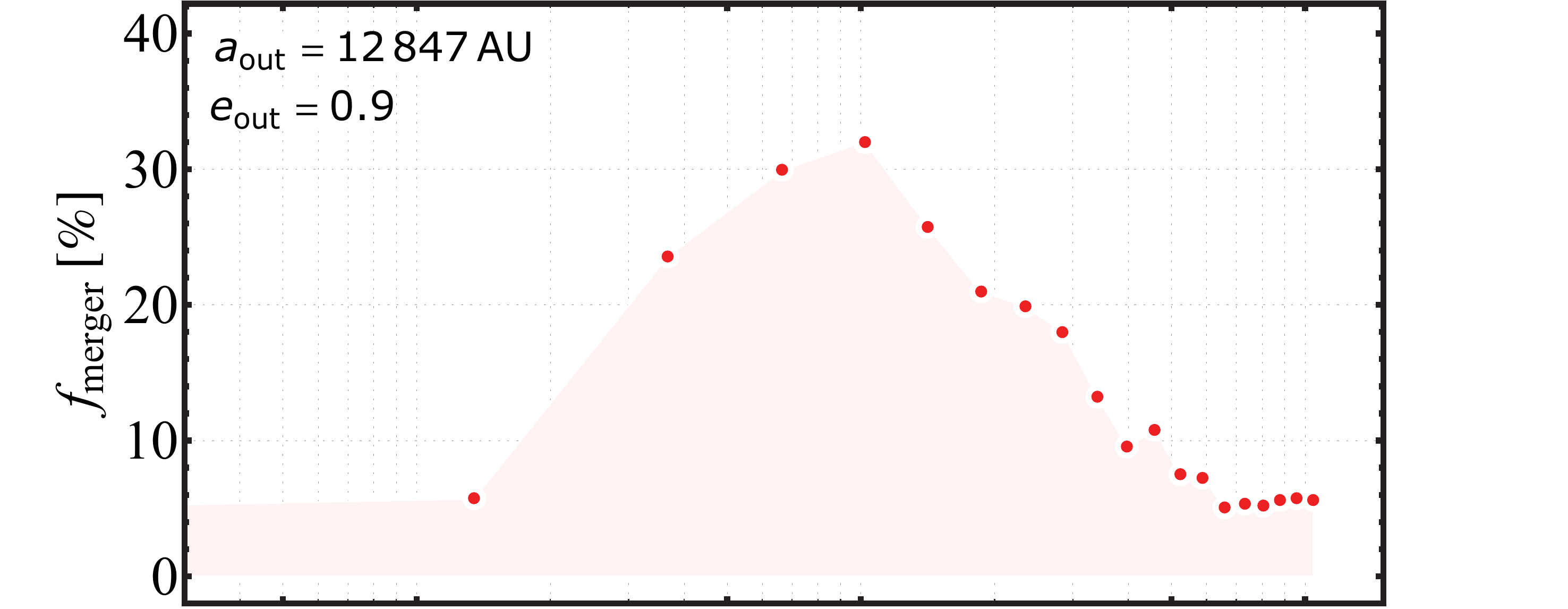}\\
\includegraphics[width=8cm]{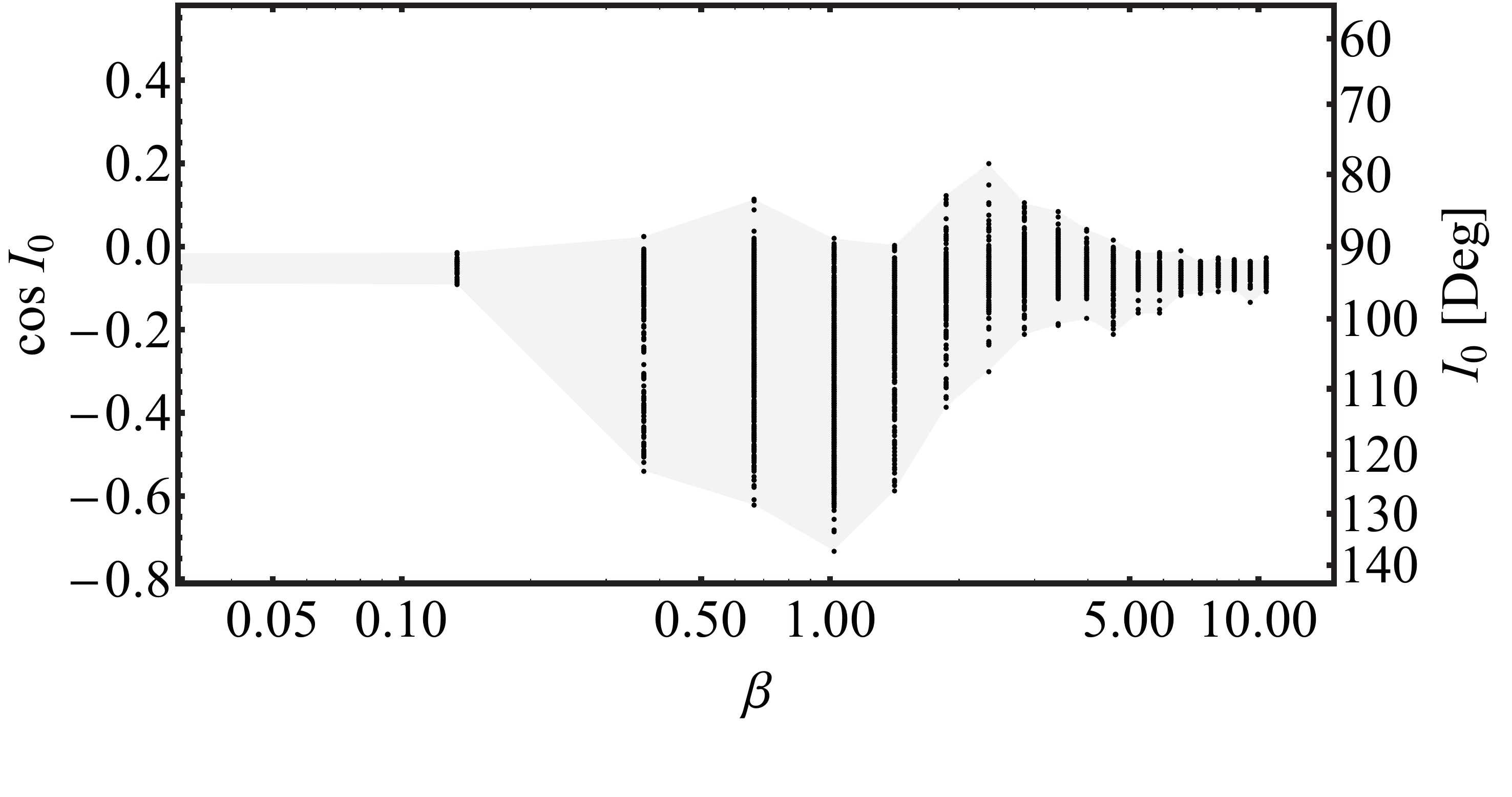}&
\includegraphics[width=8cm]{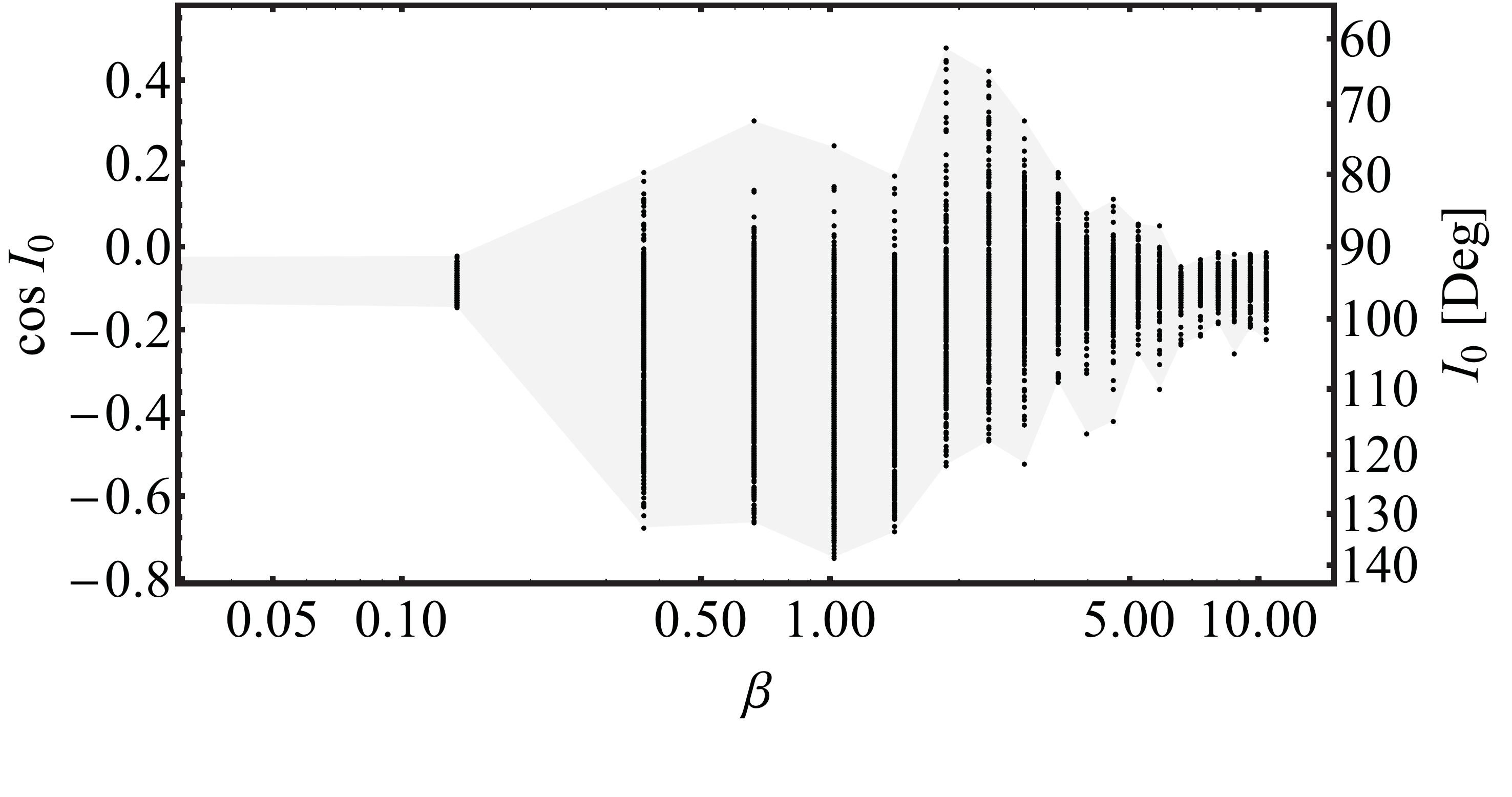}\\
\end{tabular}
\caption{Merger fraction and merger window as a function of
the dimensionless parameter $\beta$ (Equation \ref{eq:beta}) for different values of $a_\OUT$ and $e_\OUT$.
The system parameters are $m_1=30M_\odot$, $m_2=20M_\odot$, $a_0=100\au$,
$m_3=m_4=15M_\odot$, and $e_0=e_{2,0}=0.001$.
All four panels have the same $\bar{a}_{\OUT,\eff}=5.6$ (Equation \ref{eq:aout bar}) and $\omega_{\OUT}$ is initialized randomly in $(0, 2\pi)$.
In each panel, the bottom plot shows the merger window with
each dot representing a successful merger within 10 Gyrs;
the top plot shows the merger fraction
from the mergers shown in the bottom plot.
The shaded areas highlight the boundaries of the resonance parameter regimes.
}
\label{fig:merger window with Beta 1}
\end{figure*}

In the lower panels of Figure \ref{fig:merger window},
the parameters of the system (with $a_2=81\au$) gives $\beta\simeq1$.
So we indeed see that the width of LK window for extreme eccentricity excitation is significantly
enhanced due to the presence of the tertiary binary.
Note that the eccentricity of the inner binary can undergo excursions to more extreme values
than the analytical prediction of $e_\li$.
Also, when the outer binary is eccentric ($e_\OUT=0.6$), the octupole effect comes into play,
and the LK window is further extended (although slightly).
Overall, Figure \ref{fig:merger window} shows that the orbital properties of the second binary play a more important role
compared to the octupole terms in exciting eccentricity of the first inner binary and largely determine the LK window.

\section{Merger Window and Merger Fraction}
\label{sec 4}

In this section, we study the LK oscillations including gravitational radiation for binary-binary systems.
We evolve all the three binaries (using double-averaged secular equations presented in Section \ref{sec 2}).
To emphasize the role of the tertiary binary, we do not include the GW emission in the second binary and
focus on the merger window of the first inner binary
(i.e., the initial inclination $I_0$ that gives mergers in less than $10^{10}$ yrs).
Note that similar behaviors are expected when GW radiation in the second binary is considered
(if the second binary lies in the eccentricity excitation window).
In this case, the total merger rate is simply doubled.

First consider the examples shown in Figure \ref{fig:merger window}.
The upper two panels (with $a_2=1\au$, so that the second binary behaves like a single mass)
correspond to the result already found in \citet{Liu-ApJ}:
the inner binary can merge within $10^{10}$ yrs only if its eccentricity is excited
to sufficiently large value, and the merger window increases as the octupole effect (measured by $\varepsilon_\oct$)
becomes stronger. Note that for $\varepsilon_\oct=0$ and $(1-e_\m)\ll1$, the merger window can be determined analytically:
the merger time is given by
\be\label{eq:fitting formula}
T_\mathrm{m}\simeq T_\mathrm{m,0}(1-e_\m^2)^3
\ee
to a good approximation
(see Equation 48 of \citet{Liu-ApJ} and regime of validity of this equation), with $T_\mathrm{m,0}$ given by Equation (\ref{eq:Tmerger}).
Combining Equations (\ref{eq:EMAX}) and (\ref{eq:fitting formula}) and setting
$T_\mathrm{m}=10^{10}$ yrs, the upper and lower boundaries of the merger window, $I_{0,\merger}^\pm$,
can be obtained.

The lower panels of Figure \ref{fig:merger window} show that for $\beta\simeq1$, as a direct consequence
of the widened LK eccentricity excitation window, the binary merger window also significantly widens
compared to the case with small $a_2$ (or $\beta\ll1$).

In order to systematically explore how the merger window and merger fraction vary for different binary-binary parameters,
we carry out calculations for different values of $\beta$ by changing $a_2$.
Due to the large uncertainties about the stellar quadrupole populations and their properties \citep[see][]{Sana},
we are not trying to perform the calculations including full population synthesis.
Instead, we follow the fiducial BH binaries in Figure \ref{fig:merger window} and survey all possible values of $a_2$ and $a_\OUT$.
We set $m_3=m_4=15M_\odot$, but note that
the mass of the second binary $m_{34}$ is less relevant since it can be rescaled by Equation (\ref{eq:aout eff})
(i.e., the combination of $a_{\OUT,\eff}$ and $m_{34}$ determines the perturbation strength).
For a given $a_\OUT$ and $e_\OUT$, the semimajor axis of the second binary must
satisfy the stability criterion of \citet{Mardling}:
\be\label{eq:stability}
\frac{a_\OUT}{a_2}>2.8\bigg(1+\frac{m_{12}}{m_{34}}\bigg)^{2/5}\frac{(1+e_\OUT)^{2/5}}{(1-e_\OUT)^{6/5}}\bigg(1-\frac{0.3 I_{2,0}}{180^\circ}\bigg).
\ee

Figure \ref{fig:merger window with Beta 1} shows the
dependence of the merger fraction and merger window on $\beta$ for several different
semimajor axis of the outer binary, where $a_\OUT=5600\au$ ($e_\OUT=0$),
$a_\OUT=5870\au$ ($e_\OUT=0.3$), $a_\OUT=7000\au$ ($e_\OUT=0.6$) and $a_\OUT=12847\au$ ($e_\OUT=0.9$),
all given $\bar{a}_{\OUT,\eff}=5.6$ (Equation \ref{eq:aout bar})
and satisfy with the double-averaged secular approximation (Equation \ref{eq:DA condition}).
We see that the merger window indeed is much wider for $\beta\simeq0.3 - 3$.
This range is somewhat larger when the octupole effect ($\varepsilon_\oct$) increases.
Note that the initial mutual inclinations for successful mergers inside the merger window are
not uniformly distributed. This is because
the overlap of resonances from both binary-binary interactions \citep[e.g.,][]{Hamers and Lai 2017} and
octupole terms \citep[e.g.,][]{Lithwick 2011,Li chaos} together induces
chaos of the systems with intermediate $\beta$.

To calculate the merger fraction, we assume that the initial inclination of
the outer binary is uniformly distributed in $\cos I_0\in[-1,1]$.
As shown in Figure \ref{fig:merger window with Beta 1},
$f_\merger$ exhibits a clear dependence on $\beta$.
The secular resonance around $\beta\simeq1$ gives the the maximum $f_\merger\sim 30\%$,
which is $\sim6-30$ times larger than the cases with $\beta\ll1$ (equivalent to a ``pure" triple).
We also see that compared to the octupole contribution, the resonance
plays an more significant role in determining the merger fraction.

\begin{figure}
\centering
\begin{tabular}{cc}
\includegraphics[width=8cm]{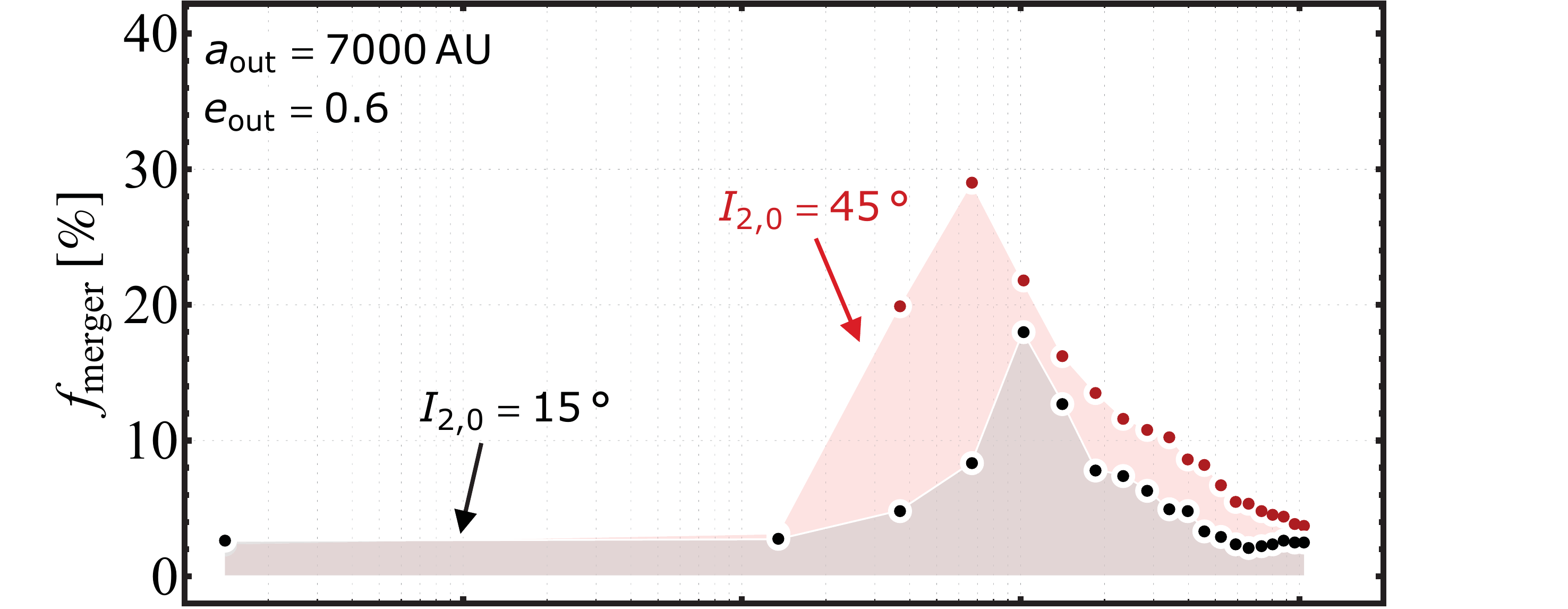}\\
\includegraphics[width=8cm]{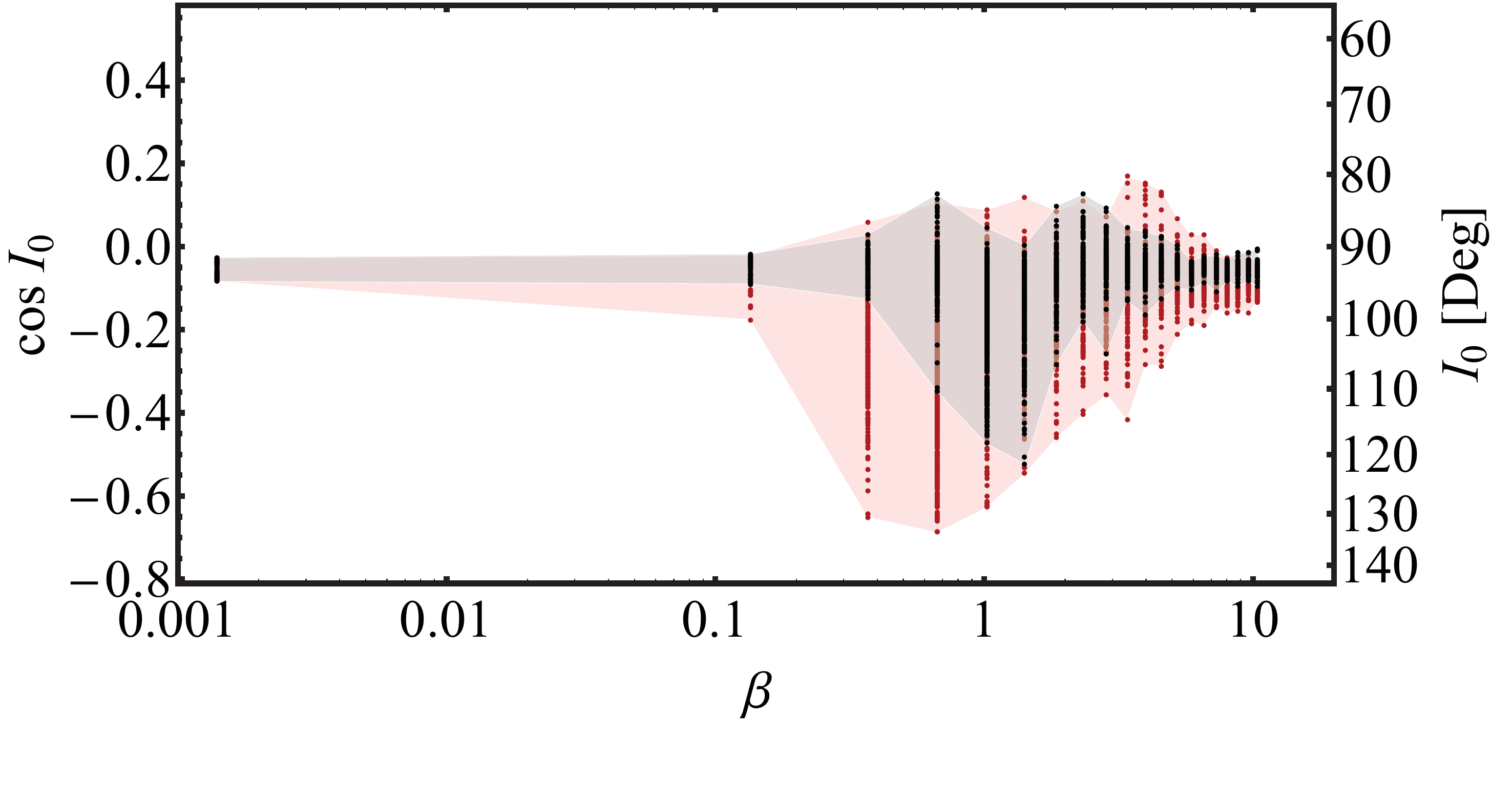}
\end{tabular}
\caption{Similar to Figure \ref{fig:merger window with Beta 1} (the $a_\OUT=7000$ AU, $e_\OUT=0.6$ case),
except for the initial $I_{2,0}=15^\circ$, $45^\circ$.
The $\beta$ value is evaluated using the initial $I_{2,0}$ (see Equation \ref{eq:beta}).
The black dots correspond to $I_{2,0}=15^\circ$ and the red dots correspond to $I_{2,0}=45^\circ$.
}
\label{fig:merger window with Beta 2}
\end{figure}

Equation (\ref{eq:beta}) indicates that $\beta$ has a dependence on $I_2$.
In the calculations shown above (Figures \ref{fig:merger window}-\ref{fig:merger window with Beta 1}),
the angular momentum vector of the second binary $\mathbf{L}_2$ is always
placed initially at $30^\circ$ with respect to $\mathbf{L}_\OUT$.
In Figure \ref{fig:merger window with Beta 2},
we set the initial $I_{2,0}$ to $15^\circ$ and $45^\circ$,
and all other parameters are sampled identically to the case of $e_\OUT=0.6$ depicted in Figure \ref{fig:merger window with Beta 1}.
The different results for $I_{2,0}=15^\circ$ and $45^\circ$ arise from the fact that
$I_2$ varies in time in the case of $I_{2,0}=45^\circ$, giving rise to
time-dependent $\beta$. Also, the amplitude of nodal precession
of the outer binary (i.e., the angle between $\mathbf{L}_\OUT$ and $\mathbf{L}_\tot$)
for the two cases are different, and this difference can affect the LK oscillations of the first inner binary
\citep[see][]{Hamers and Lai 2017}.

\begin{figure*}
\begin{centering}
\begin{tabular}{cccc}
\includegraphics[width=8cm]{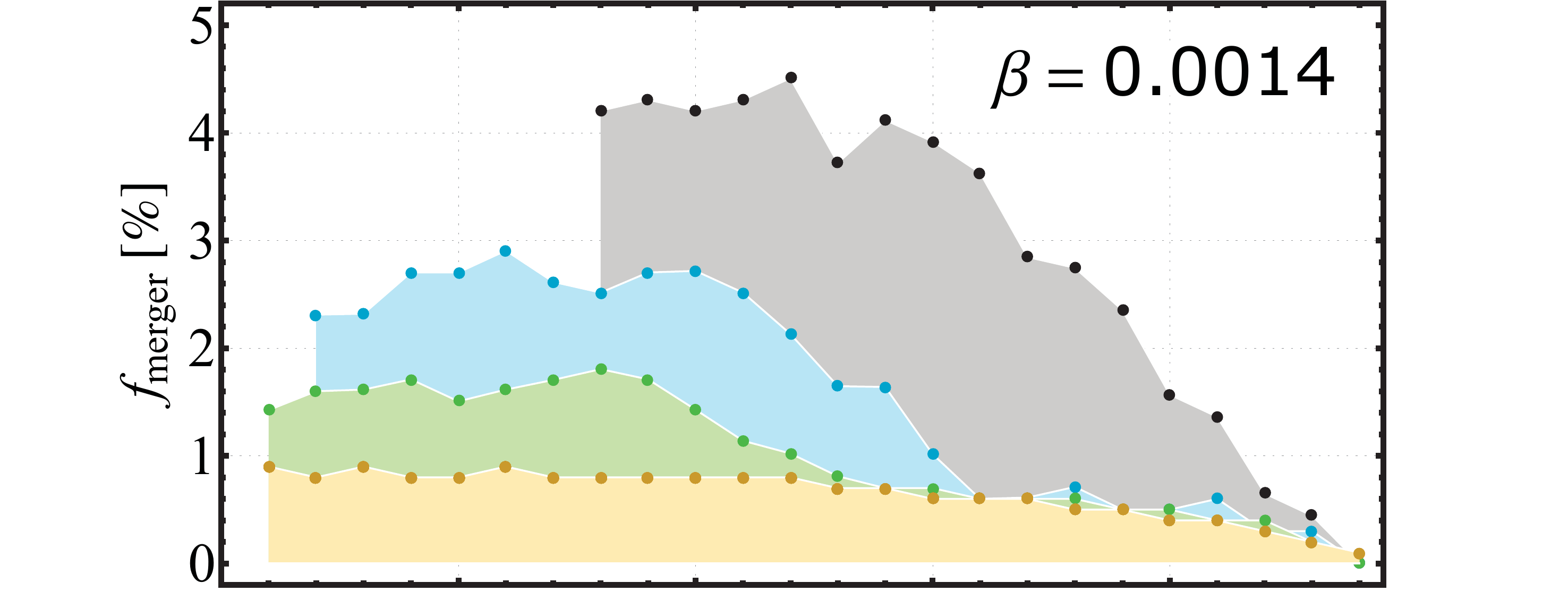}&
\includegraphics[width=8cm]{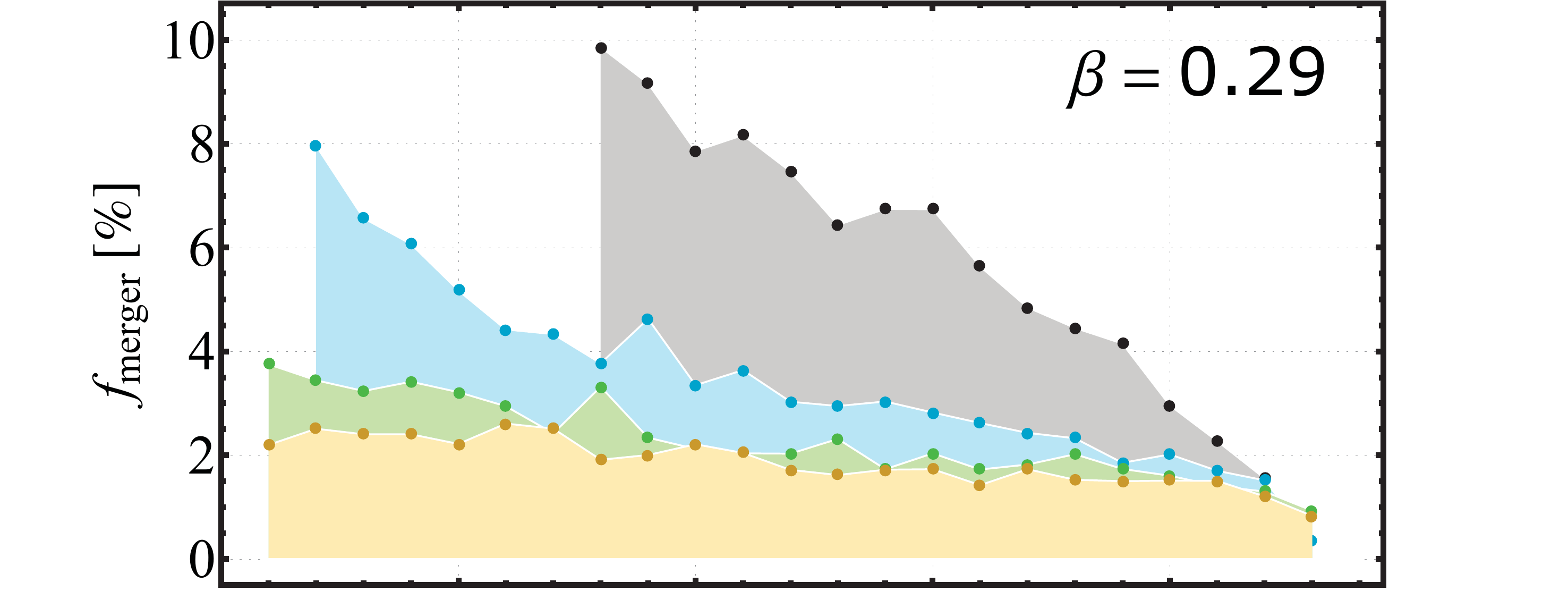}\\
\includegraphics[width=8cm]{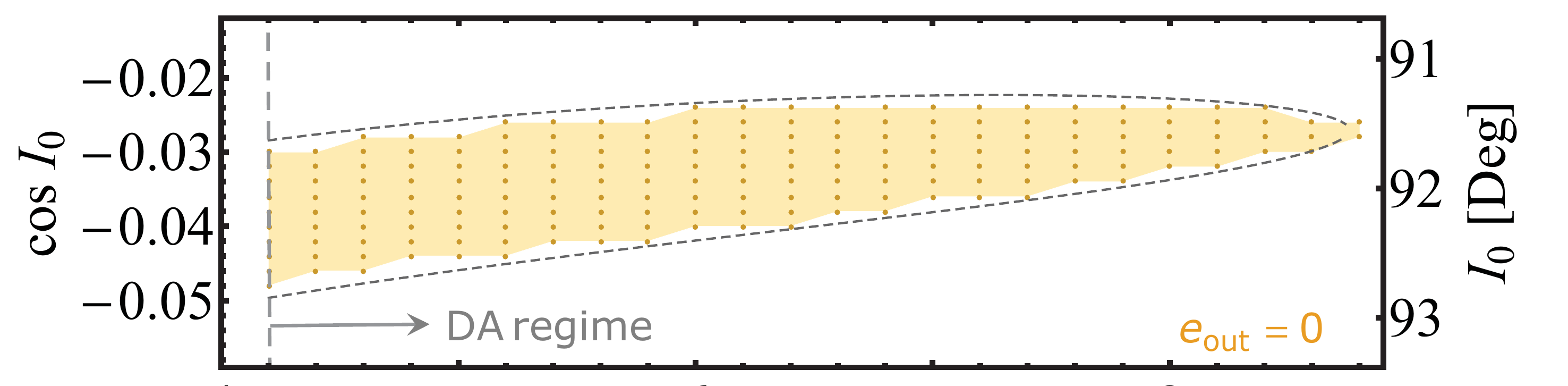}&
\includegraphics[width=8cm]{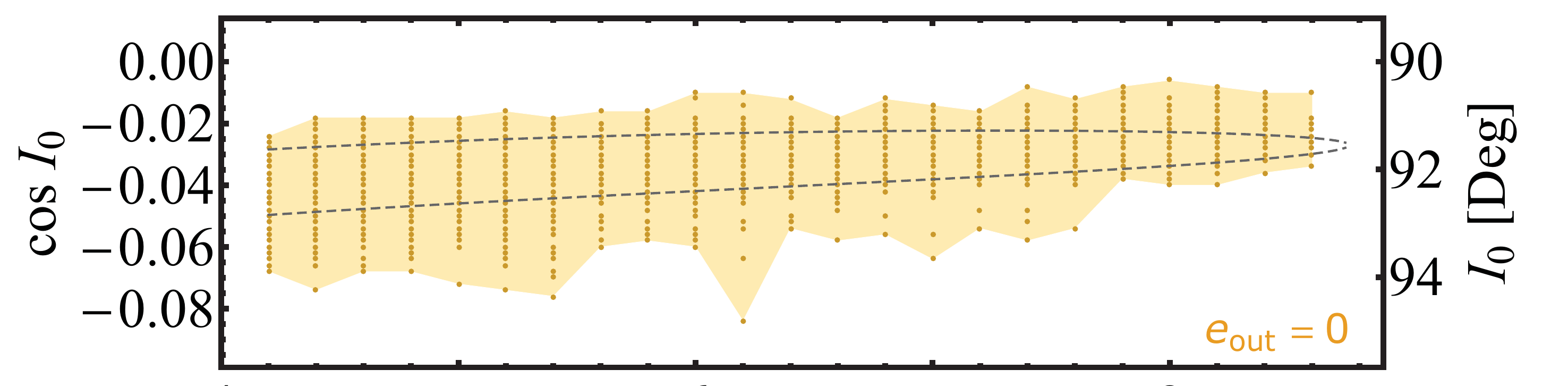}\\
\includegraphics[width=8cm]{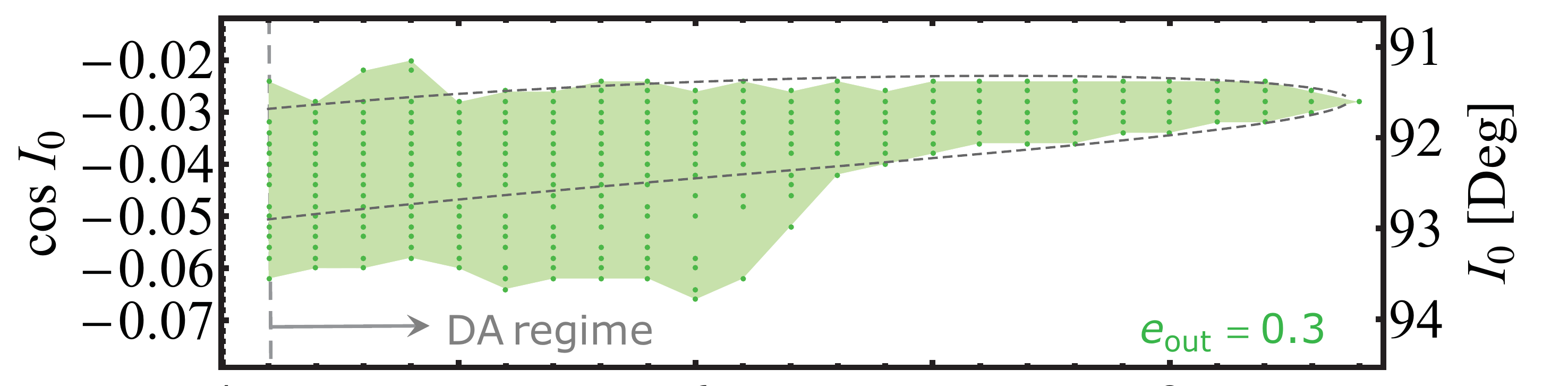}&
\includegraphics[width=8cm]{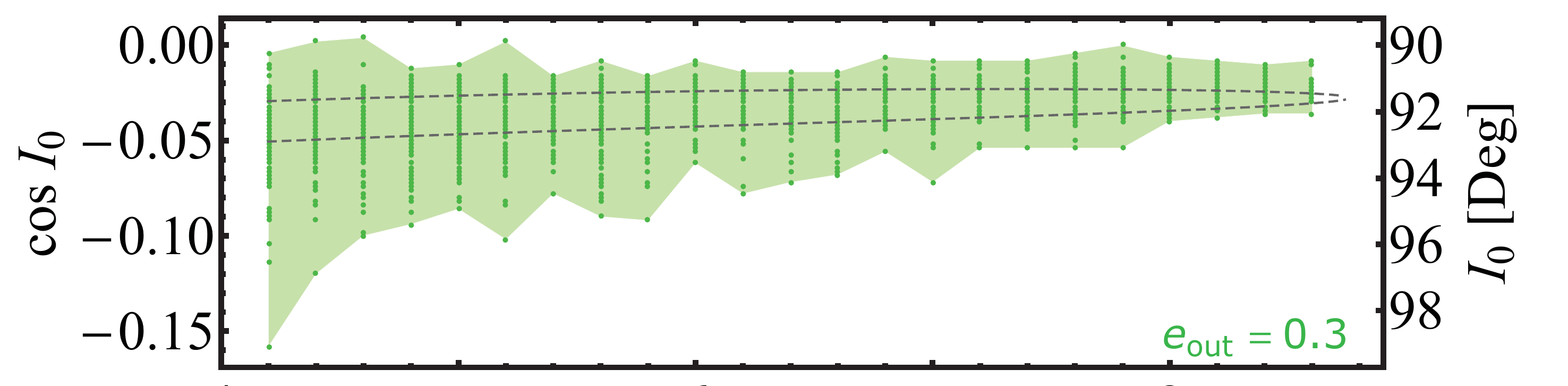}\\
\includegraphics[width=8cm]{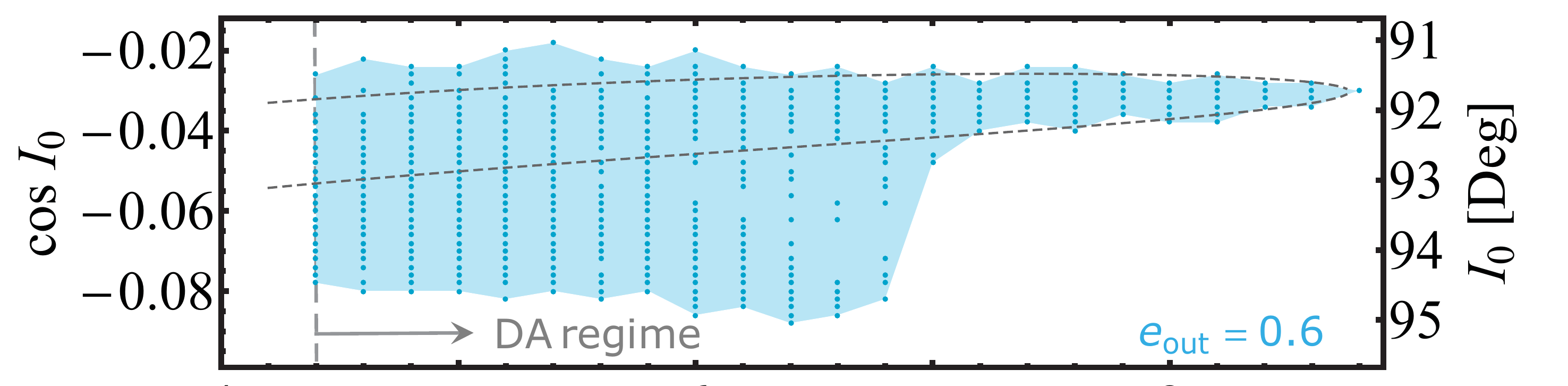}&
\includegraphics[width=8cm]{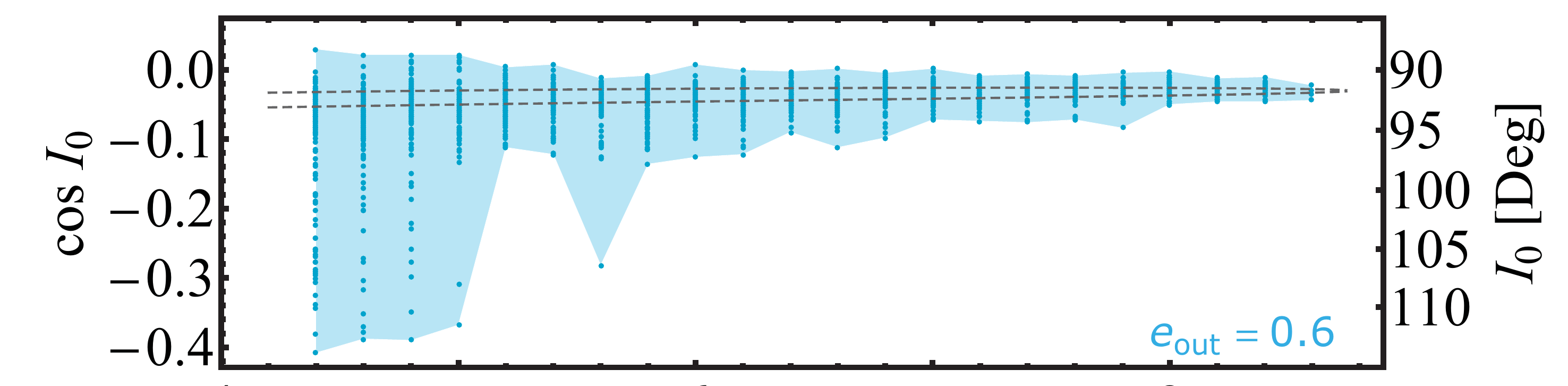}\\
\includegraphics[width=8cm]{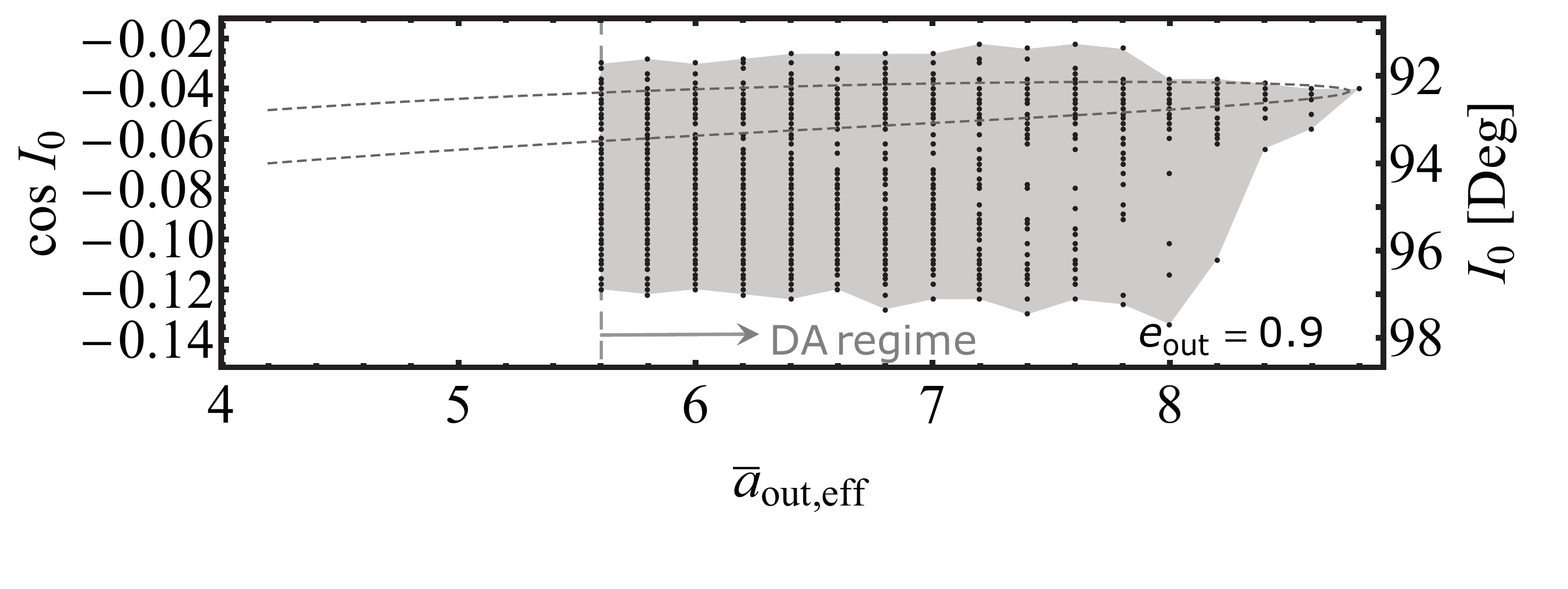}&
\includegraphics[width=8cm]{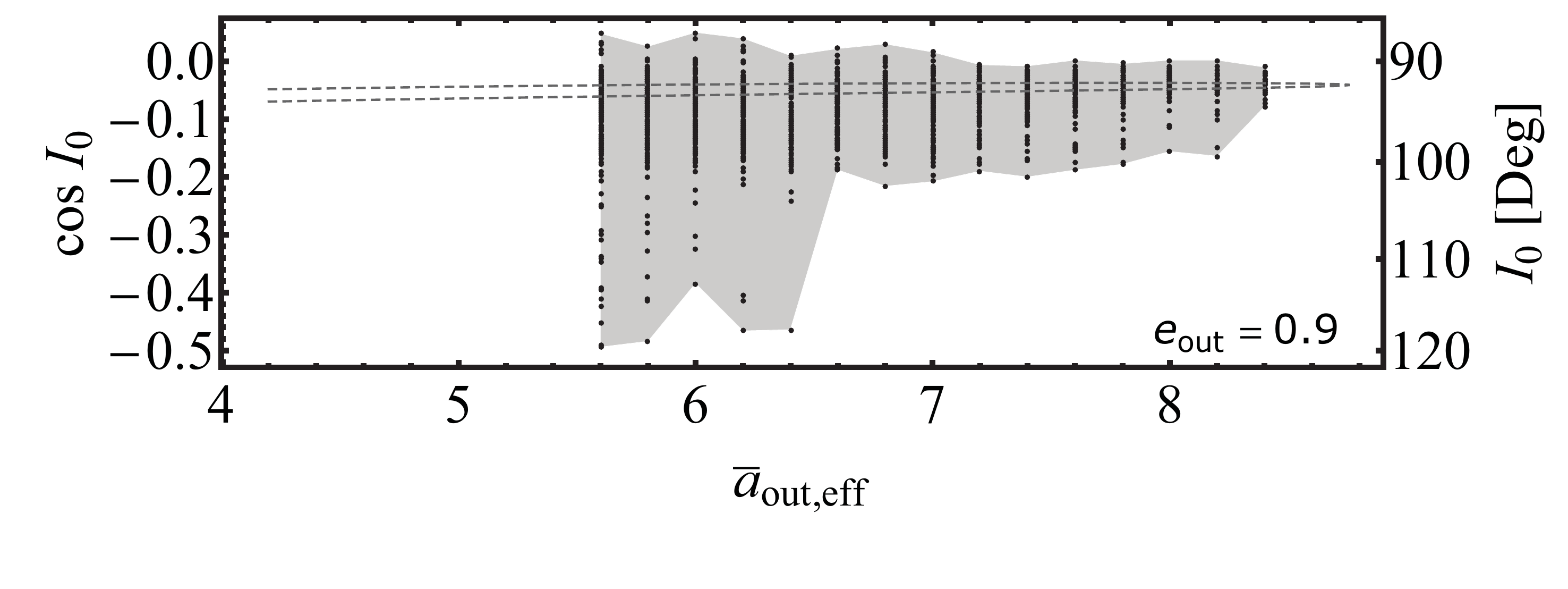}\\
\end{tabular}
\caption{Merger window and merger fraction as a function of
the effective semi-major axis of outer binary $\bar{a}_{\OUT,\eff}$ (Equation \ref{eq:aout bar})
for $\beta=0, 0.29$ (corresponding to $a_2=1\au$ and $35\au$).
Other parameters are the same as the example in Figure \ref{fig:merger window with Beta 1}
In all examples, we assume a fixed $I_{2,0}=30^\circ$.
In each case, each dot (in the bottom four panels)
represents a successful merger event within $10^{10}$ yrs.
Note that merger events can have an irregular distribution as a function of $\cos I_0$,
because of the chaotic behavior introduced by the octupole terms and binary-binary interactions.
Also note that we only consider the range of $\bar{a}_{\OUT,\eff}$ such that double-averaged secular equations are valid
(see Equation \ref{eq:DA condition}). Different color represents the numerical results from different $e_\OUT$ and
the shaded areas highlight the boundaries of the merger windows.
}
\label{fig:merger fraction1}
\end{centering}
\end{figure*}

\begin{figure*}
\begin{centering}
\begin{tabular}{cccc}
\includegraphics[width=8cm]{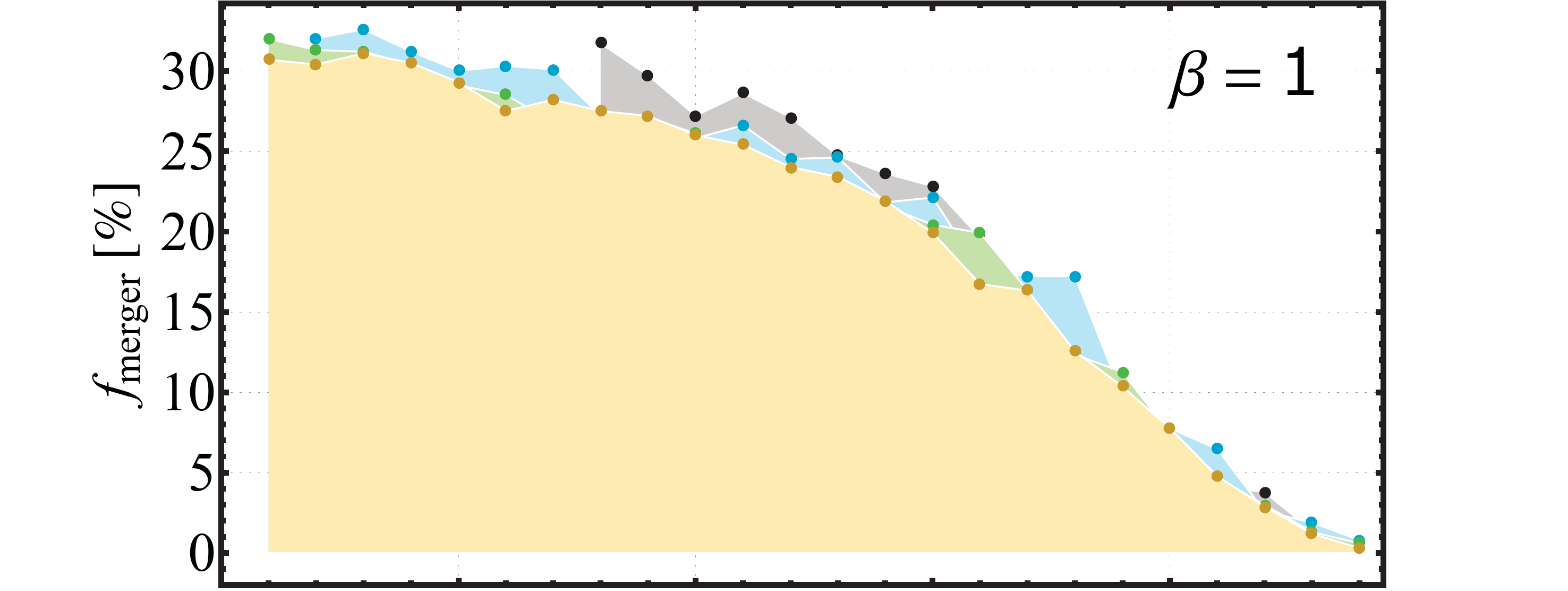}&
\includegraphics[width=8cm]{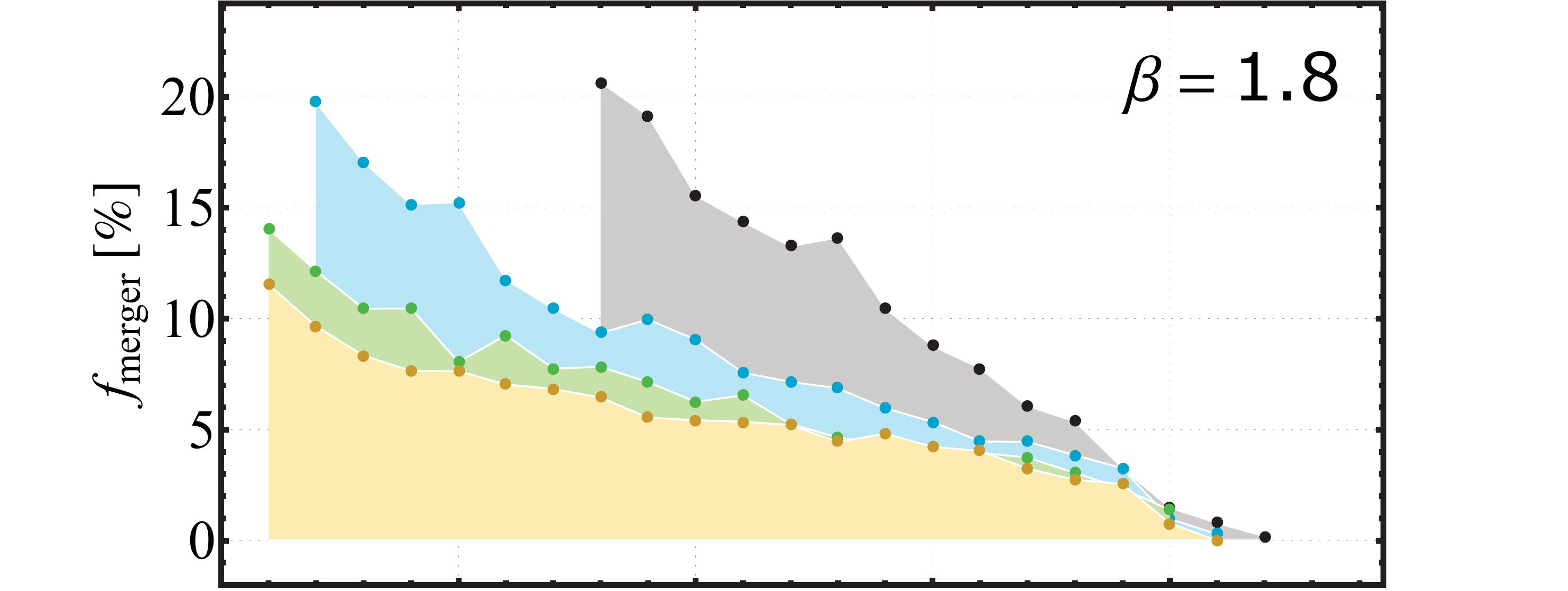}\\
\includegraphics[width=8cm]{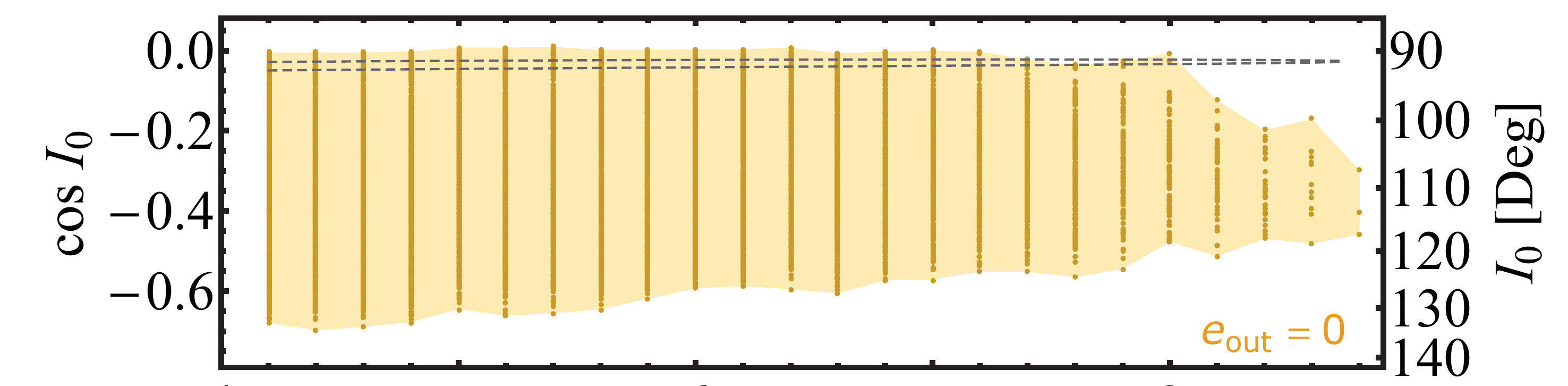}&
\includegraphics[width=8cm]{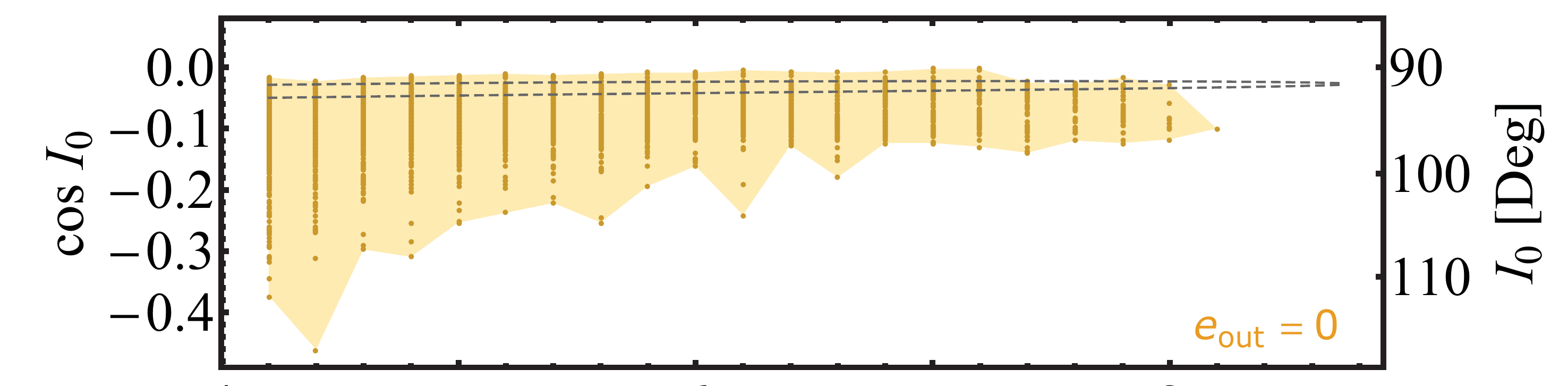}\\
\includegraphics[width=8cm]{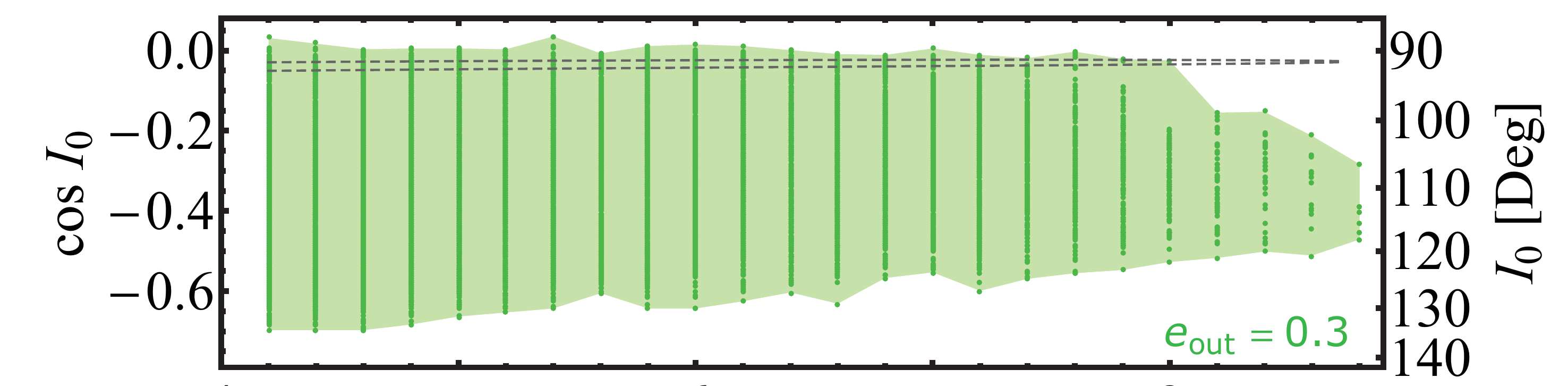}&
\includegraphics[width=8cm]{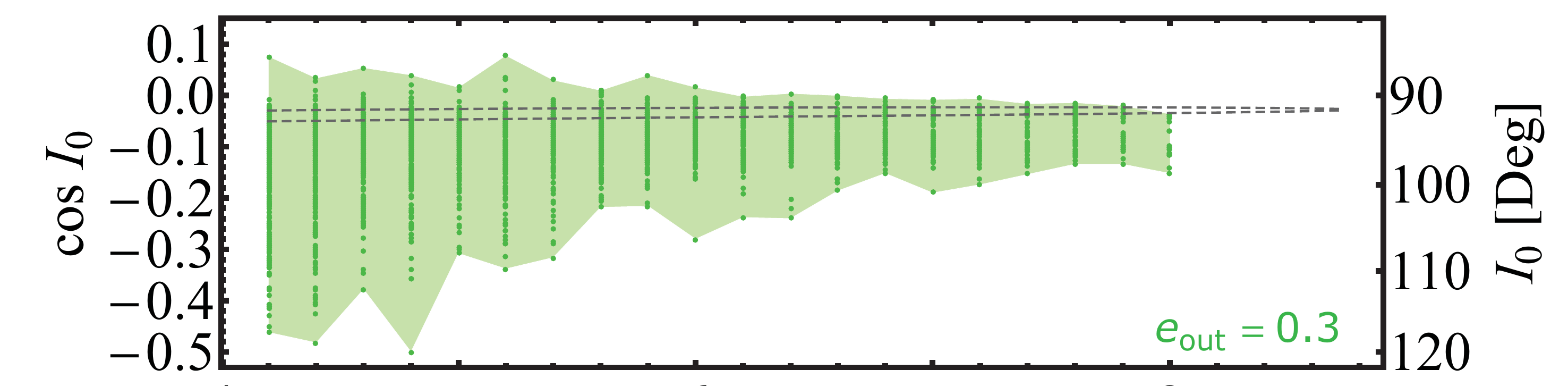}\\
\includegraphics[width=8cm]{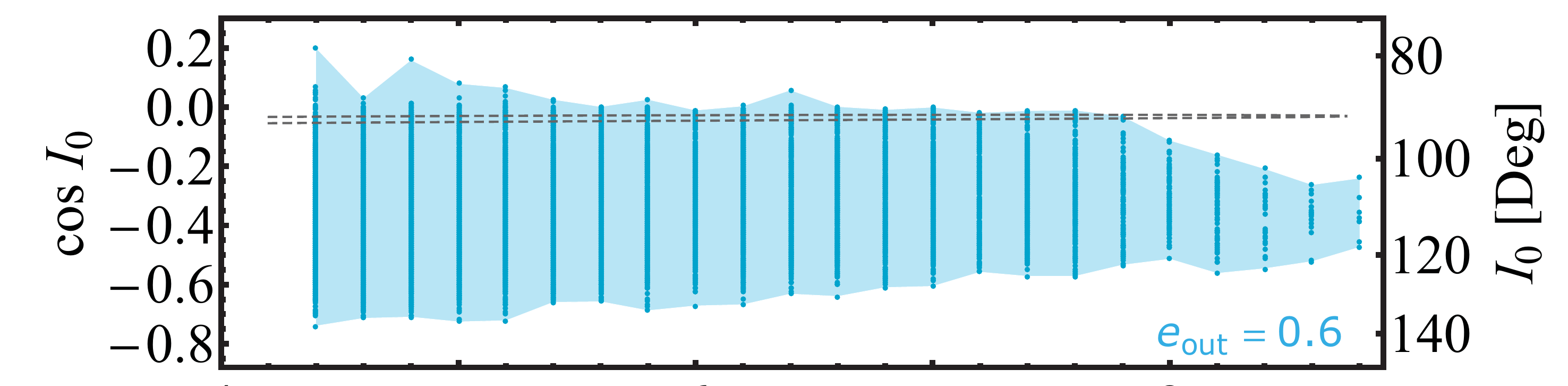}&
\includegraphics[width=8cm]{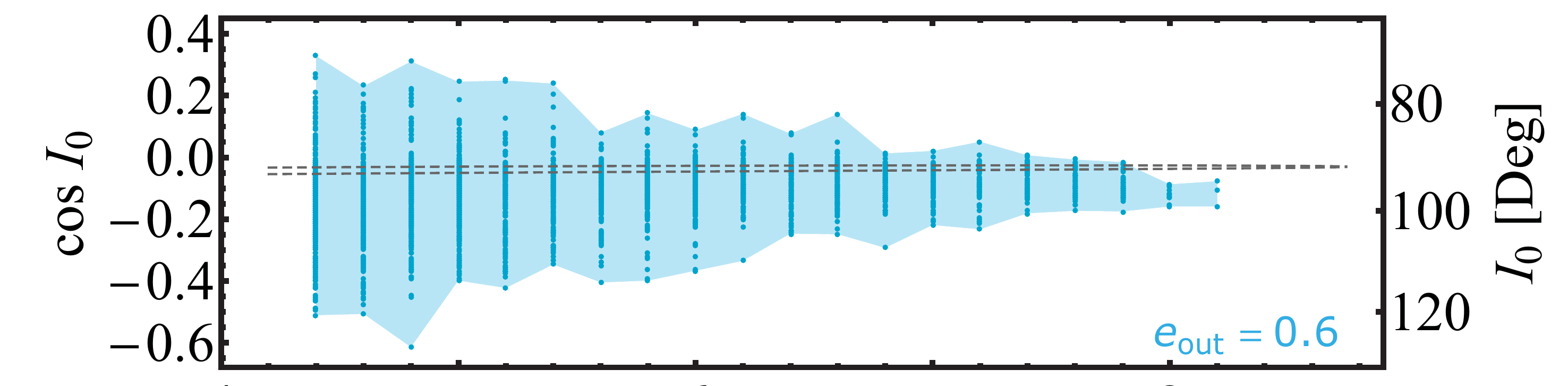}\\
\includegraphics[width=8cm]{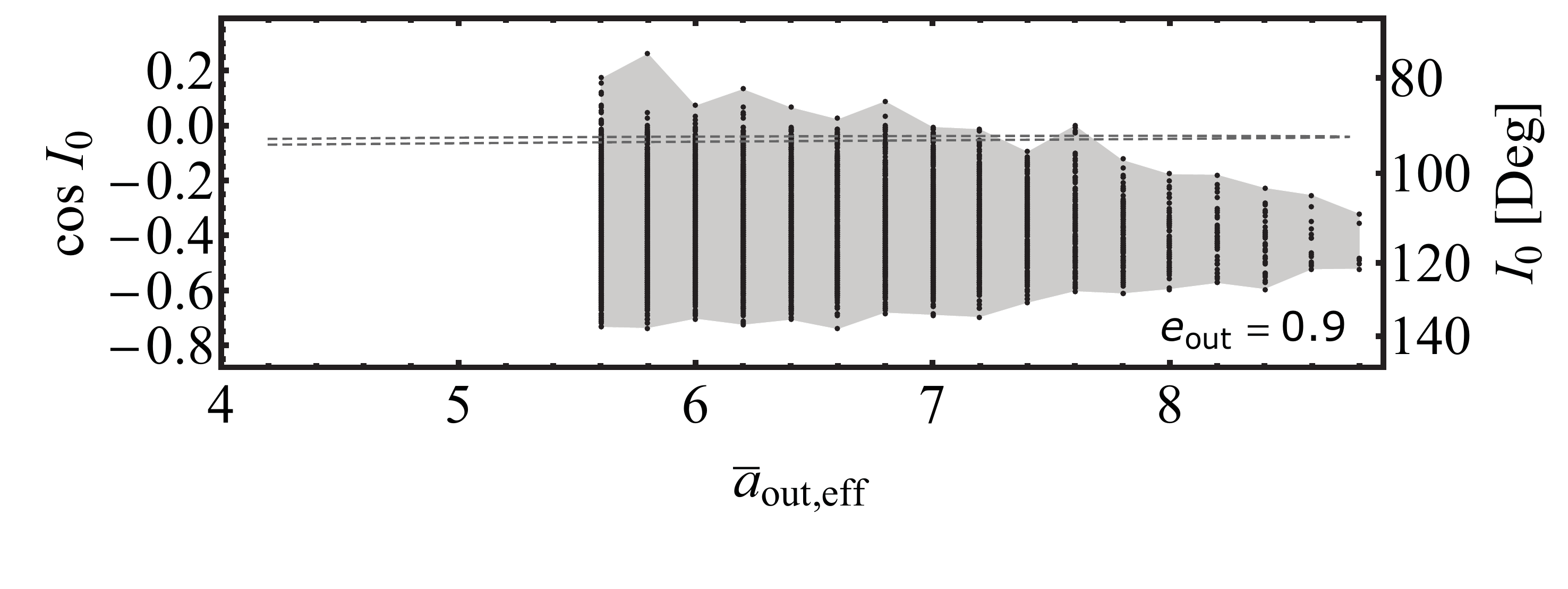}&
\includegraphics[width=8cm]{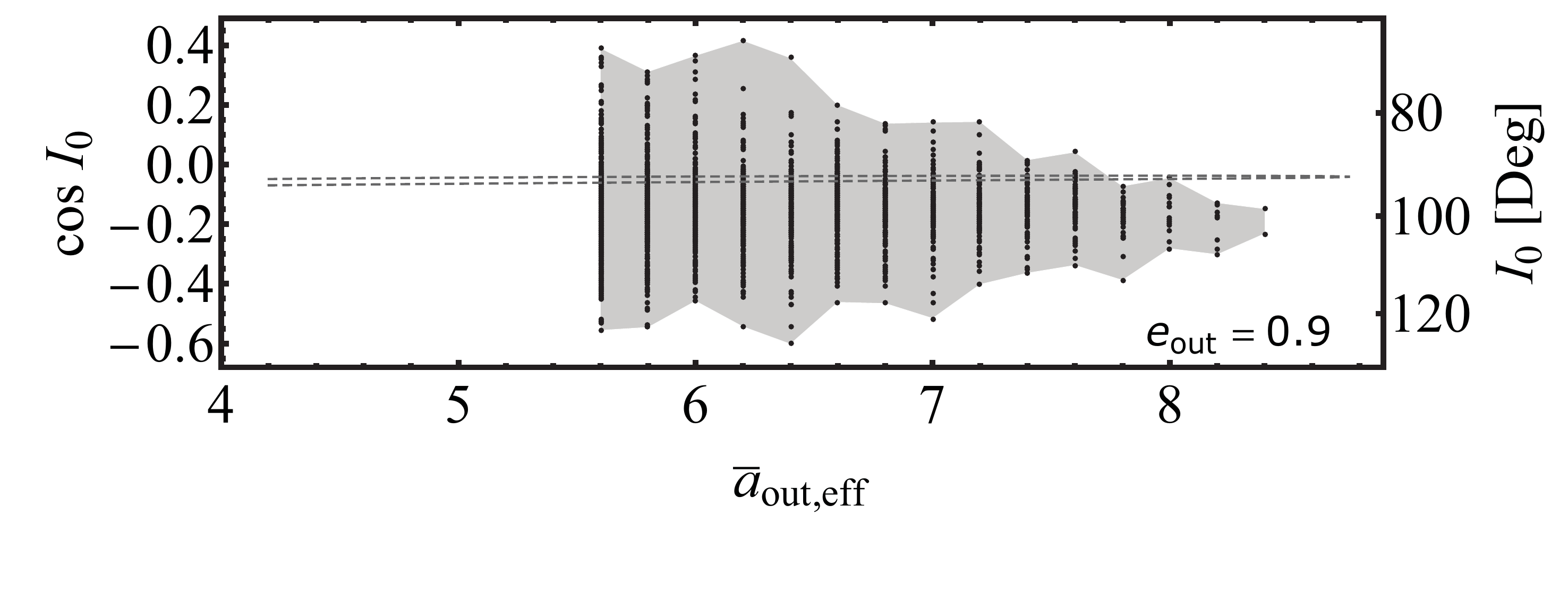}
\end{tabular}
\caption{Same as Figure \ref{fig:merger fraction1}, except for $\beta=1, 1.8$ (corresponding to $a_2=81\au$ and $121\au$).
}
\label{fig:merger fraction2}
\end{centering}
\end{figure*}

To illustrate how the merger window and merger fraction depend on the properties of the outer binary,
Figures \ref{fig:merger fraction1}-\ref{fig:merger fraction2} show our results
as a function of $\bar{a}_{\OUT,\eff}$ for several values of $\beta$.
This is similar to Figure \ref{fig:merger window with Beta 1}, but shown in a different way.
When $\beta\ll1$, for a given $e_\OUT$, the merger window shows an general trend of widening as
$\bar{a}_{\OUT,\eff}$ decreases. Note that for $e_\OUT\simeq0$, the merger window (the dashed curve in each panel)
and merger fraction can be
obtained analytically using Equations (\ref{eq:EMAX}) and (\ref{eq:fitting formula}) \citep[see Equations 51, 53 and 54 of ][]{Liu-ApJ}.
For the same value of $\bar{a}_{\OUT,\eff}$ (thus the same quadrupole effect),
the merger window and merger fraction can be different for different $e_\OUT$.
In general, the larger the eccentricity $e_\OUT$, the stronger the octupole effect, and the wider the window.
The merger fraction ranges from $\sim1\%$ (for $e_\OUT\simeq0$) to a few $\%$
(for $e_\OUT=0.9$). Note that for some values of $\bar{a}_{\OUT,\eff}$,
the irregular distribution of merger events inside the merger window is evident; this results from
the chaotic behaviors of the octupole-level LK oscillations
(see also the examples in Figure \ref{fig:merger window}, particulary the $e_\OUT=0.6$ case).

For $0.3\lesssim\beta\lesssim3$, the merger window and merger fraction are significantly larger for all
values of $e_\OUT$. At $\beta\simeq1$, different values of $e_\OUT$ give the similar $f_\merger$ for each $\bar{a}_{\OUT,\eff}$.
The secular resonance enhances $f_\merger$ to tens of percent.

\begin{figure}
\begin{centering}
\begin{tabular}{cccc}
\includegraphics[width=7.5cm]{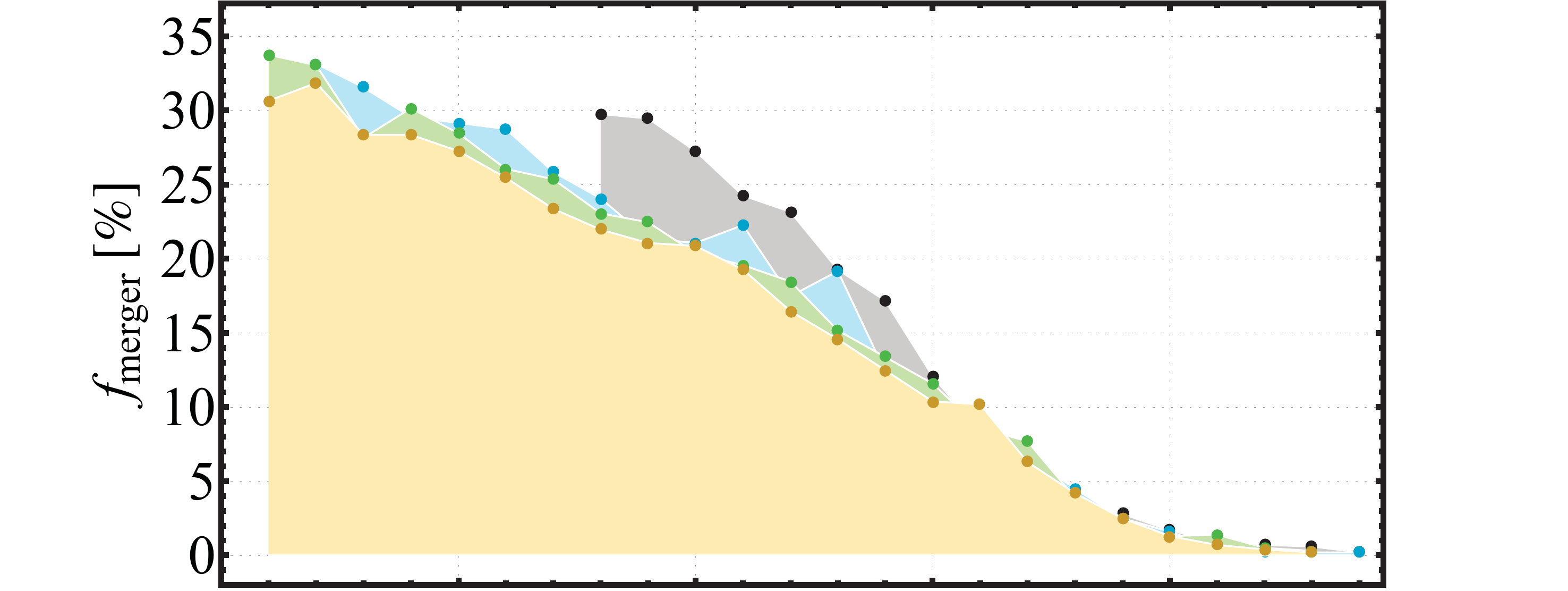}\\
\includegraphics[width=7.5cm]{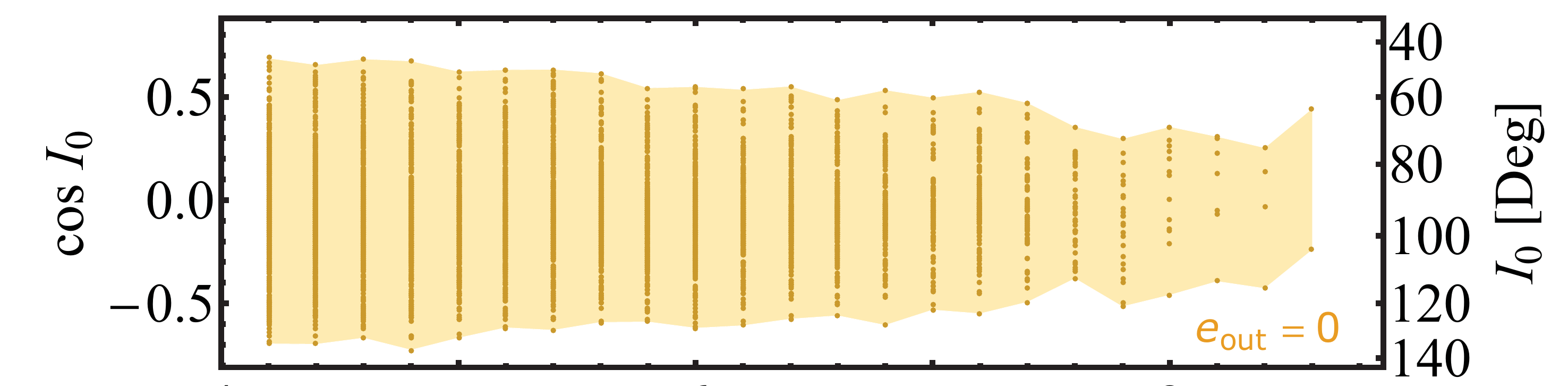}\\
\includegraphics[width=7.5cm]{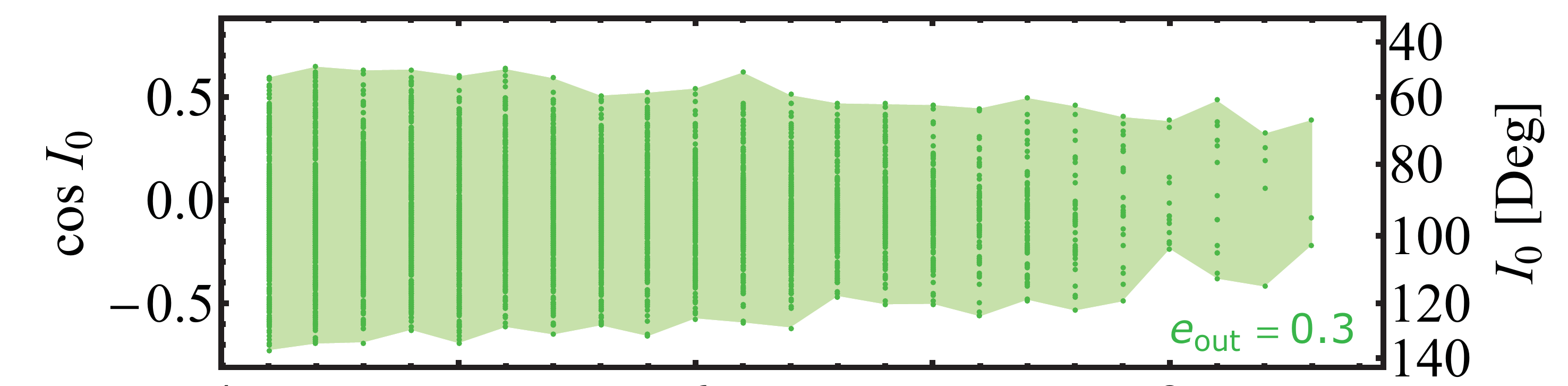}\\
\includegraphics[width=7.5cm]{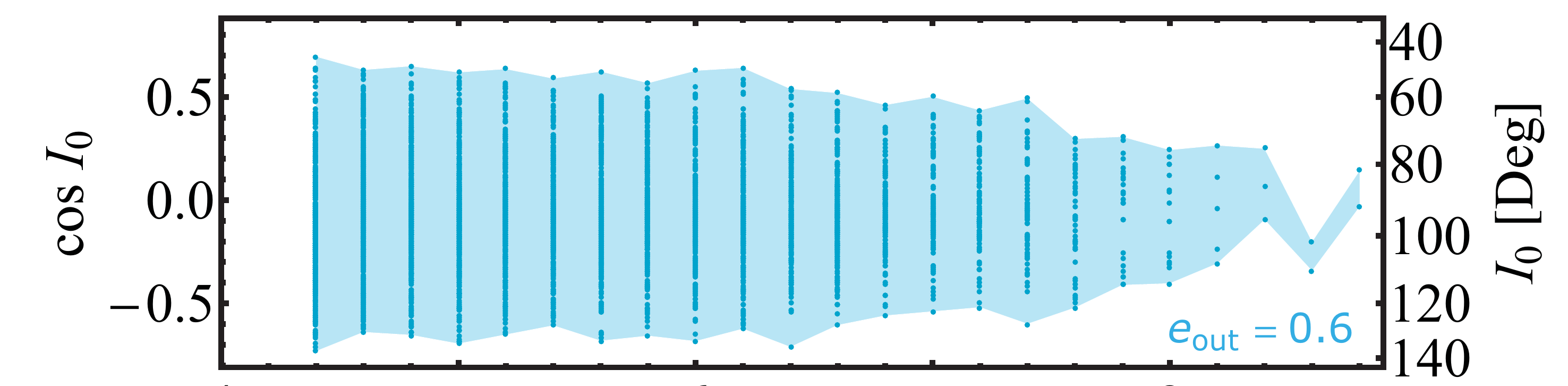}\\
\includegraphics[width=7.5cm]{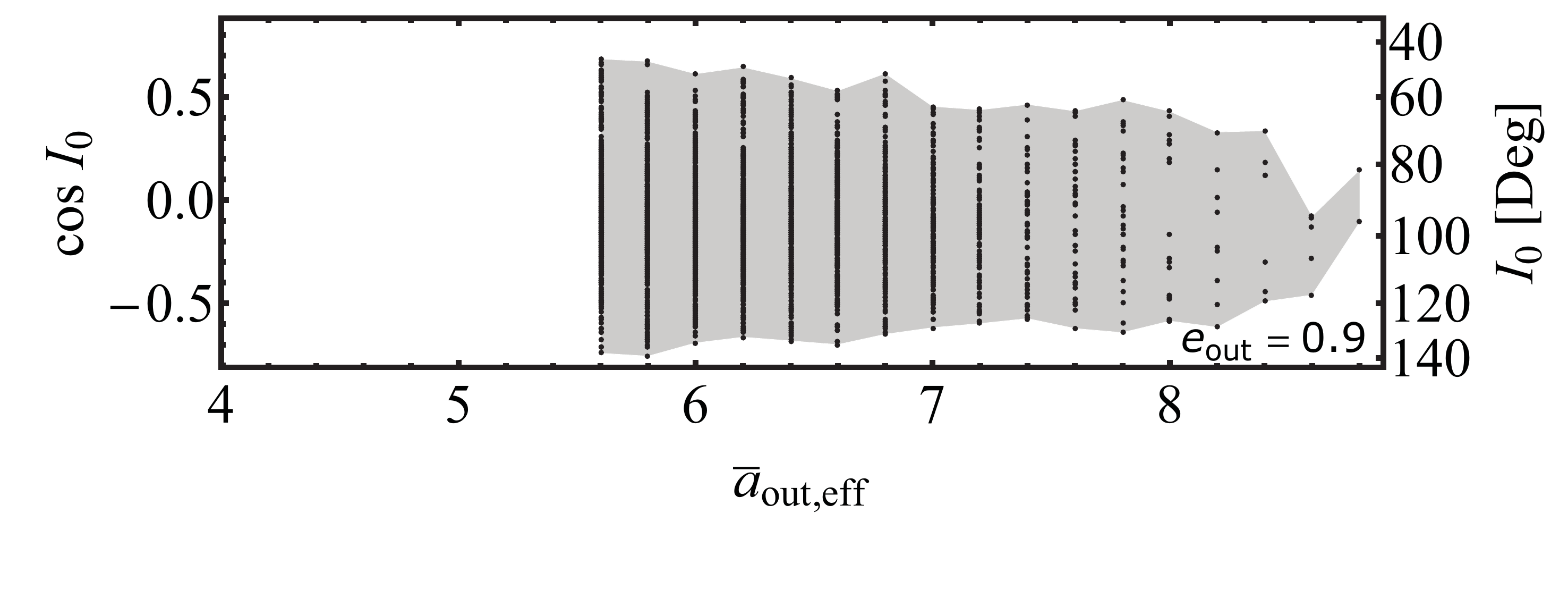}
\end{tabular}
\caption{Similar to the case of $a_2=81\au$ in Figure \ref{fig:merger fraction2},
except that the initial $\cos I_{2,0}$ is distributed uniformly in $\cos I_{2,0}\in[-1,1]$.
}
\label{fig:merger fraction 2}
\end{centering}
\end{figure}

If the orbital plane of the second inner binary has initially random orientation,
LK oscillations in the second binary become possible, and the merger window and merger fraction can be changed.
We show an example in Figure \ref{fig:merger fraction 2} for the case of $a_2=81\au$,
and the absolute value of $\beta\propto \cos I_2$ is in the range from $0$ to $1.18$.
We see that the merger fraction for the random $\cos I_{2,0}$ case is similar to
the $\beta\sim1$ case depicted in Figures \ref{fig:merger fraction1}-\ref{fig:merger fraction2}. However, unlike Figures \ref{fig:merger fraction1}-\ref{fig:merger fraction2}, where
the window lies in the retrograde regime ($\cos I_0<0$), in Figure \ref{fig:merger fraction 2} a large
fraction of mergers occurs in the prograde regime ($\cos I_0>0$).

Note that the merger fractions presented above are based on the fiducial
inner BH binary parameters ($m_1=30M_\odot$, $m_2=20M_\odot$, $a_0=100\au$).
If we start with a closer binary or consider moderately hierarchical systems,
where the double-averaged secular approximation may break down, the merger fraction can be
even higher \citep[see][]{Liu-ApJ}.

\begin{figure}
\begin{centering}
\begin{tabular}{cccc}
\includegraphics[width=8.5cm]{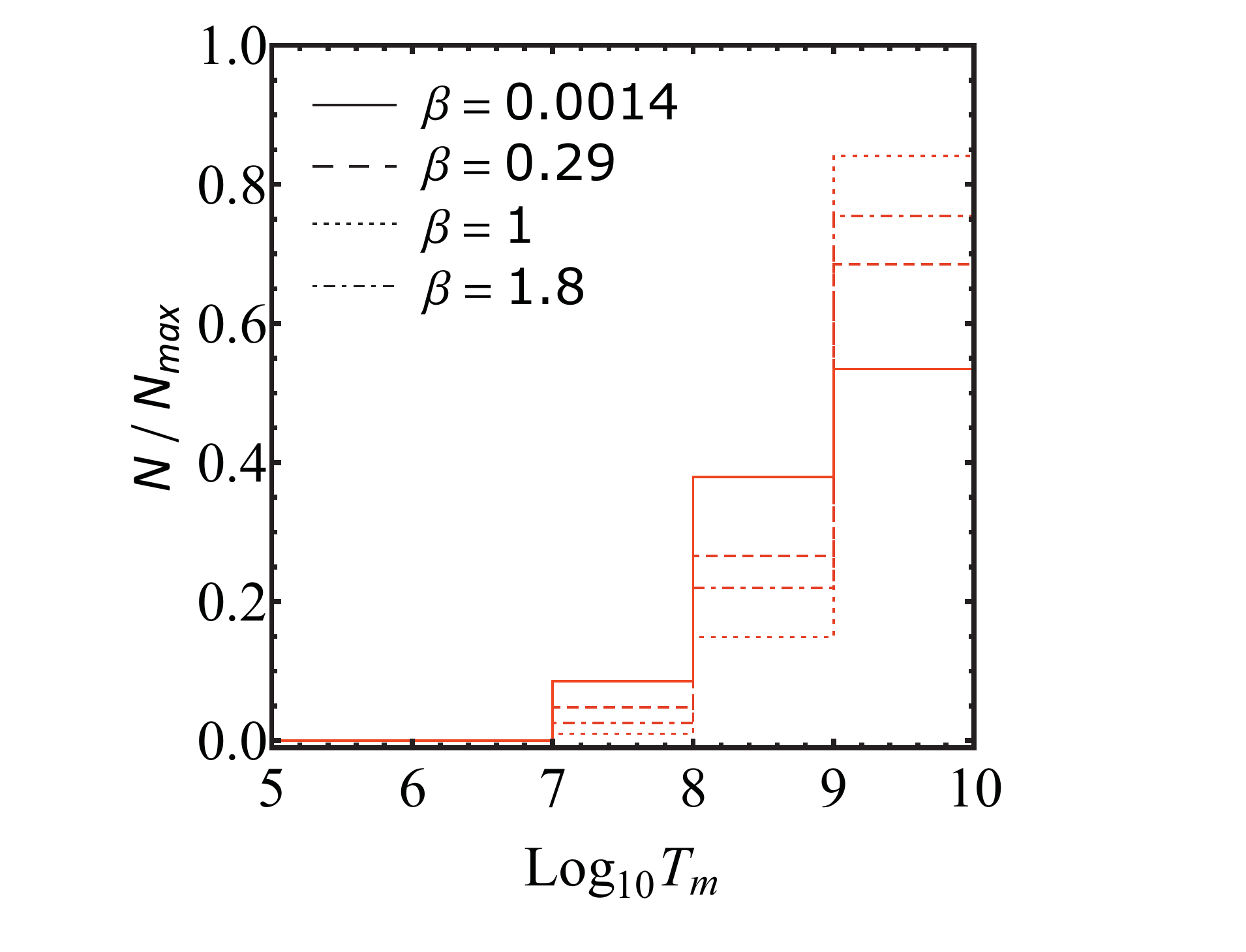}
\end{tabular}
\caption{The distribution of merger time (in years) normalized by the number of mergers for different values of $\beta$.
Here we only include merging systems with $\bar{a}_{\OUT,\eff}\in[5.6, 8.8]$ in Figures \ref{fig:merger fraction1}-\ref{fig:merger fraction2}.
For each $\beta$, the number of mergers is 986 ($\beta=0.0014$), 1913 ($\beta=0.29$), 11359 ($\beta=1$), 3374 ($\beta=1.8$), respectively.
}
\label{fig:merger Time}
\end{centering}
\end{figure}

Having studied the role of binaries, we now
summarize the distribution of the merger time for the merging systems studied in Figures \ref{fig:merger fraction1}-\ref{fig:merger fraction2}.
we consider systems with $\bar{a}_{\OUT,\eff}\in[5.6, 8.8]$,
and assume that the eccentricity of the
tertiary companion has a uniform distribution in $e_\OUT$
(i.e., $e_\OUT=0, 0.3, 0.6, 0.9$ are equally probable),
and the initial mutual inclination is randomly distributed (uniform in $\cos I_0$).
Figure \ref{fig:merger Time} shows the result for four values of $\beta$.
We see that most systems take long time to merge
(with $T_\mathrm{m}\sim10^9-10^{10}$ yrs).
In particular, a larger fraction of the systems with $\beta\simeq1$ merge
with $T_\mathrm{m}>10^9$ yrs,
compared to those with $\beta\ll1$.
This is because when $\bar{a}_{\OUT,\eff}\gtrsim5.6$,
the merger window for systems with $\beta\simeq1$ is always larger than the other systems ($\beta=0.0014, 0.29, 1.8$),
providing more merger events even with the same quadrupole perturbation (same $\bar{a}_{\OUT,\eff}$).

\section{Summary and Discussion}
\label{sec 5}

In this paper, we have studied the mergers of binary BHs induced
by the gravitational interaction with tertiary binaries.
The binary-binary system is evolved in time using the octupole-level secular equations of motion,
taking account of the post-Newtonian effect and gravitational radiation.
We examine the dependence of the eccentricity excitation of the BH binary
on the orbital properties of the tertiary binary.
When the precession timescale of the outer orbit driven by the tertiary binary is
comparable to the Lidov-Kozai oscillation time of the BH binary
($\beta\simeq1$; Equation \ref{eq:beta}),
the LK inclination window for $e$-excitation is enhanced drastically,
leading to more BH mergers compared to the standard triple
(``binary + perturber") systems (see Figure \ref{fig:merger window}).

By conducting a series of numerical integrations,
we quantify the role of tertiary binaries in determining the BH merger
windows and merger fractions. We find that the orbital properties of the external binary
(especially the semi-major axis $a_2$) play a more important role
in producing large merger
fractions compared to the octupole effect (i.e., eccentric outer orbit).
When $\beta\ll1$ or $\beta\gg1$, the merger windows are similar as in the standard triples,
with the merger fraction less than a few $\%$
(see Figure \ref{fig:merger window with Beta 1}). This gives the lower limit of the merger fraction
in the binary-binary interaction channel. However,
for systems with $\beta \sim 0.3-3$,
the merger fraction increases to $\gtrsim 10\%$, peaking at $\sim 30\%$,
depending on the parameters of the outer orbits (see Figures \ref{fig:merger fraction1}-\ref{fig:merger fraction2}).
This places the upper limit to the BH merger fraction due to the presence of tertiary binaries.
Note that our numerical results are based on the fiducial parameters of first BH binary
($m_{1,2}\simeq 20M_\odot-30M_\odot$ and initial $a_0\sim100$ AU).
However, our analysis is not restricted to any specific systems and can be safely extended to
other configurations of system parameters. For example,
for a relatively compact BH binary, a higher merger fraction induced by the binary-binary interactions could be expected.
In triple-driven scenario, similar results can be found in \citep[][]{Liu-ApJ}.

In this paper we have focused on BH binaries in bound orbits around
another binaries. To determine the global BH binary merger rate from
such binary-binary channel, we would need to start from a population
of main-sequence stellar quadruples, follow them through stellar
evolution and BH formation, and eventually to eccentricity excitation
and binary mergers. Such calculation is highly uncertain, and is
beyond the scope of this paper. Recent population studies of BH
binary mergers from field stellar triples gave a global merger rate of
a few per Gpc$^3$ per year, which is within the low end of the
observed BH merger rate determined by LIGO \citep[][]{Silsbee and Tremaine 2017,Antonini 2017}.
The multiplicity fraction of
high-mass main-sequence stars is quite high (as large as $90\%$), with each star
having more than 2 companions on average, suggesting that the stellar quadruple
fraction is not much smaller than the stellar triple fraction \citep[][]{Sana}.
With our finding that the merger fraction of quadruple systems is
about 10 times larger than that of triple systems, we conclude that
dynamically driven BH mergers in binary-binary systems may be more important
than those produced in triple systems, and contribute appreciably to the BH merger events observed by LIGO/VIRGO.

\section{Acknowledgments}

This work is supported in part by the NSF grant AST-1715246 and NASA grant NNX14AP31G.
BL is also supported in part by grants from
NSFC (No. 11703068 and No. 11661161012). This work made use of the High
Performance Computing Resource in the Core Facility for Advanced
Research Computing at Shanghai Astronomical Observatory.

\label{lastpage}


\begin{thebibliography}{}

\bibitem[\protect\citeauthoryear{Abbott et al.}{2016a}]{Abbott 2016a} Abbott B.~P., et al., 2016a, PhRvL, 116, 061102

\bibitem[\protect\citeauthoryear{Abbott et al.}{2016b}]{Abbott 2016b} Abbott B.~P., et al., 2016b, PhRvL, 116, 241103

\bibitem[\protect\citeauthoryear{Abbott et al.}{2017a}]{Abbott 2017a} Abbott B.~P., et al., 2017a, PhRvL, 118, 221101

\bibitem[\protect\citeauthoryear{Abbott et al.}{2017b}]{Abbott 2017b} Abbott B.~P., et al., 2017b, ApJ, 851, L35

\bibitem[\protect\citeauthoryear{Abbott et al.}{2017c}]{Abbott 2017c} Abbott B.~P., et al., 2017c, PhRvL, 119, 141101

\bibitem[\protect\citeauthoryear{Abbott et al.}{2017d}]{Abbott 2017d} Abbott B.~P., et al., 2017d, PhRvL, 119, 161101

\bibitem[\protect\citeauthoryear{Anderson et al.}{2017}]{Anderson et al 2017} Anderson K.~R., Lai D., Storch N.~I., 2017, MNRAS, 467, 3066

\bibitem[\protect\citeauthoryear{Antonini \& Perets}{2012}]{Antonini 2012} Antonini F., Perets H.~B., 2012, ApJ, 757, 27

\bibitem[\protect\citeauthoryear{Antonini et al.}{2014}]{Antonini 2014} Antonini F., Murray N., Mikkola S., 2014, ApJ, 781, 45

\bibitem[\protect\citeauthoryear{Antonini \& Rasio}{2016}]{Antonini and Rasio 2016} Antonini F., Rasio F.~A., 2016, ApJ, 831, 187

\bibitem[\protect\citeauthoryear{Antonini et al.}{2017}]{Antonini 2017} Antonini F., Toonen S., Hamers A.~S., 2017, ApJ, 841, 77

\bibitem[\protect\citeauthoryear{Banerjee et al.}{2010}]{Banerjee 2010} Banerjee S., Baumgardt H., Kroupa P., 2010, MNRAS, 402, 371

\bibitem[\protect\citeauthoryear{Belczynski et al.}{2010}]{Belczynski 2010} Belczynski K., Dominik M., Bulik T., O'Shaughnessy R., Fryer C., Holz D.~E., 2010, ApJ, 715, L138

\bibitem[\protect\citeauthoryear{Belczynski et al.}{2016}]{Belczynski 2016} Belczynski K., Holz D.~E., Bulik T., O'Shaughnessy R., 2016, Natur, 534, 512

\bibitem[\protect\citeauthoryear{Chatterjee et al.}{2017}]{Chatterjee 2017} Chatterjee S., Rodriguez C.~L., Kalogera V., Rasio F.~A., 2017, ApJ, 836, L26

\bibitem[\protect\citeauthoryear{Dominik et al.}{2012}]{Dominik 2012} Dominik M., Belczynski K., Fryer C., Holz D.~E., Berti E., Bulik T., Mandel I., O'Shaughnessy R., 2012, ApJ, 759, 52

\bibitem[\protect\citeauthoryear{Dominik et al.}{2013}]{Dominik 2013} Dominik M., Belczynski K., Fryer C., Holz D.~E., Berti E., Bulik T., Mandel I., O'Shaughnessy R., 2013, ApJ, 779, 72

\bibitem[\protect\citeauthoryear{Dominik et al.}{2015}]{Dominik 2015} Dominik M., et al., 2015, ApJ, 806, 263

\bibitem[\protect\citeauthoryear{Downing et al.}{2010}]{Downing 2010} Downing J.~M.~B., Benacquista M.~J., Giersz M., Spurzem R., 2010, MNRAS, 407, 1946

\bibitem[\protect\citeauthoryear{Fang et al.}{2018}]{Fang 2018} Fang X., Thompson T.~A., Hirata C.~M., 2018, MNRAS, 476, 4234

\bibitem[\protect\citeauthoryear{Hamers \& Portegies Zwart}{2016}]{Hamers 2016} Hamers A.~S., Portegies Zwart S.~F., 2016, MNRAS, 459, 2827

\bibitem[\protect\citeauthoryear{Hamers \& Lai}{2017}]{Hamers and Lai 2017} Hamers A.~S., Lai D., 2017, MNRAS, 470, 1657

\bibitem[\protect\citeauthoryear{Hamers}{2018a}]{Hamers (2018)} Hamers A.~S., 2018, MNRAS, 478, 620

\bibitem[\protect\citeauthoryear{Hamers et al.}{2018b}]{Hamers 2018b}
Hamers A.~S., Bar-Or B., Petrovich C., Antonini F., 2018, arXiv, arXiv:1805.10313

\bibitem[\protect\citeauthoryear{Hoang et al.}{2018}]{Hoang 2017} Hoang B.-M., Naoz S., Kocsis B., Rasio F.~A., Dosopoulou F., 2018, ApJ, 856, 140

\bibitem[\protect\citeauthoryear{Kozai}{1962}]{Kozai} Kozai Y., 1962, AJ, 67, 591

\bibitem[\protect\citeauthoryear{Leigh et al.}{2018}]{Leigh 2018} Leigh N.~W.~C., et al., 2018, MNRAS, 474, 5672

\bibitem[\protect\citeauthoryear{Li et al.}{2015}]{Li chaos} Li G., Naoz S., Holman M., Loeb A., 2014, ApJ, 791, 86

\bibitem[\protect\citeauthoryear{Lidov}{1962}]{Lidov} Lidov M. L., 1962, Planet. Space Sci., 9, 719

\bibitem[\protect\citeauthoryear{Lithwick et al. }{2011}]{Lithwick 2011} Lithwick Y., Naoz S., 2011, ApJ, 742, 94

\bibitem[\protect\citeauthoryear{Lipunov et al.}{1997}]{Lipunov 1997} Lipunov V.~M., Postnov K.~A., Prokhorov M.~E., 1997, AstL, 23, 492

\bibitem[\protect\citeauthoryear{Lipunov et al.}{2017}]{Lipunov 2017} Lipunov V.~M., et al., 2017, MNRAS, 465, 3656

\bibitem[\protect\citeauthoryear{Liu et al.}{2015}]{Liu et al 2015} Liu B., Mu{\~n}oz D.~J., Lai D., 2015, MNRAS, 447, 747

\bibitem[\protect\citeauthoryear{Liu \& Lai}{2018}]{Liu-ApJ} Liu B., Lai D., 2018,  ApJ, 863, 68

\bibitem[\protect\citeauthoryear{Mandel \& de Mink}{2016}]{Mandel and de Mink 2016} Mandel I., de Mink S.~E., 2016, MNRAS, 458, 2634

\bibitem[\protect\citeauthoryear{Marchant et al.}{2016}]{Marchant 2016} Marchant P., Langer N., Podsiadlowski P., Tauris T.~M., Moriya T.~J., 2016, A\&A, 588, A50

\bibitem[\protect\citeauthoryear{Mardling \& Aarseth}{2001}]{Mardling} Mardling R. A., Aarseth S.J., 2001, MNRAS, 321, 398

\bibitem[\protect\citeauthoryear{Miller \& Hamilton}{2002}]{Miller 2002} Miller M.~C., Hamilton D.~P., 2002, ApJ, 576, 894

\bibitem[\protect\citeauthoryear{Miller \& Lauburg}{2009}]{Miller 2009} Miller M.~C., Lauburg V.~M., 2009, ApJ, 692, 917

\bibitem[\protect\citeauthoryear{Naoz}{2016}]{Naoz 2016} Naoz S., 2016, ARA\&A, 54, 441

\bibitem[\protect\citeauthoryear{O'Leary et al.}{2006}]{O'Leary 2006} O'Leary R.~M., Rasio F.~A., Fregeau J.~M., Ivanova N., O'Shaughnessy R., 2006, ApJ, 637, 937

\bibitem[\protect\citeauthoryear{O'Leary et al.}{2009}]{O'Leary 2009} O'Leary R.~M., Kocsis B., Loeb A., 2009, MNRAS, 395, 2127

\bibitem[\protect\citeauthoryear{Pejcha et al.}{2013}]{Pejcha quadple} Pejcha O., Antognini J.~M., Shappee B.~J., Thompson T.~A., 2013, MNRAS, 435, 943

\bibitem[\protect\citeauthoryear{Peters}{1964}]{Peters 1964} Peters P.~C., 1964, PhRv, 136, 1224

\bibitem[\protect\citeauthoryear{Petrovich}{2015}]{Petrovich 2015} Petrovich C., 2015, ApJ, 799, 27

\bibitem[\protect\citeauthoryear{Petrovich \& Antonini}{2017}]{Petrovich 2017} Petrovich C., Antonini F., 2017, ApJ, 846, 146

\bibitem[\protect\citeauthoryear{Podsiadlowski et al.}{2003}]{Podsiadlowski 2003} Podsiadlowski P., Rappaport S., Han Z., 2003, MNRAS, 341, 385

\bibitem[\protect\citeauthoryear{Portegies Zwart \& McMillan}{2000}]{Portegies 2000} Portegies Zwart S.~F., McMillan S.~L.~W., 2000, ApJ, 528, L17

\bibitem[\protect\citeauthoryear{Rodriguez et al.}{2015}]{Rodriguez 2015} Rodriguez C.~L., Morscher M., Pattabiraman B., Chatterjee S., Haster C.-J., Rasio F.~A., 2015, PhRvL, 115, 051101

\bibitem[\protect\citeauthoryear{Sana}{2017}]{Sana} Sana H., 2017, IAUS, 329, 110

\bibitem[\protect\citeauthoryear{Samsing et al.}{2018}]{Samsing 2018} Samsing J., D'Orazio D.~J., Askar A., Giersz M., 2018, arXiv, arXiv:1802.08654

\bibitem[\protect\citeauthoryear{Seto}{2013}]{Seto PRL} Seto N., 2013, PhRvL, 111, 061106

\bibitem[\protect\citeauthoryear{Silsbee \& Tremaine}{2017}]{Silsbee and Tremaine 2017} Silsbee K., Tremaine S., 2017, ApJ, 836, 39

\bibitem[\protect\citeauthoryear{Thompson}{2011}]{Thompson 2011} Thompson T.~A., 2011, ApJ, 741, 82

\bibitem[\protect\citeauthoryear{VanLandingham et al.}{2016}]{VanLandingham 2016} VanLandingham J.~H., Miller M.~C., Hamilton D.~P., Richardson D.~C., 2016, ApJ, 828, 77

\bibitem[\protect\citeauthoryear{Vokrouhlick{\'y}}{2016}]{Vokrouhlicky quadruple} Vokrouhlick{\'y} D., 2016, MNRAS, 461, 3964

\bibitem[\protect\citeauthoryear{Wen}{2003}]{Wen 2003} Wen L., 2003, ApJ, 598, 419

\end{thebibliography}
\end{document}